%% file: asepfin1.tex
\documentclass[11pt]{article}
\usepackage{amssymb}
\font\bigsym=cmr10 scaled\magstep3
\font\dlsbig=cmmi10  scaled\magstep3
\font\dlsmed=cmmi10  scaled\magstep2
 \input prepictex
 \input pictex
 \input postpictex

\let\tsection\section
\renewcommand{\section}{\setcounter{equation}{0}\tsection}

\def\F{{\cal F}}

\def\fncomma{$^{\rm ,}$}

\def\rhob{\bar\rho}

\begin{document}
%\today
\begin{center} 

EXACT LARGE DEVIATION FUNCTIONAL OF A STATIONARY OPEN DRIVEN DIFFUSIVE 
SYSTEM: THE ASYMMETRIC EXCLUSION PROCESS

\vskip10pt

B. Derrida,\footnote{Laboratoire de Physique Statistique,
Ecole Normale Sup\'erieure, 24 rue Lhomond, 75005 Paris, France; 
email derrida@lps.ens.fr.}%
    \fncomma${}^{3}$
  J. L. Lebowitz,\footnote{Department of Mathematics,
Rutgers University, New Brunswick, NJ 08903; email lebowitz@math.rutgers.edu,
speer@math.rutgers.edu.}%
    \fncomma\footnote{Also School of Mathematics, Institute
for Advanced Study, Princeton, NJ 08540.}%
    \fncomma\footnote{Also Department of Physics, Rutgers.}
  and E. R. Speer${}^{2}$

%\today
\vskip20pt

Dedicated to Michael E. Fisher on the occasion of his seventieth birthday.
\end{center} 

\vskip20pt
\noindent {\bf Abstract}

 We consider the asymmetric exclusion process (ASEP)
in one dimension on sites $i = 1,..., N$, in contact at sites
$i=1$ and $i=N$ with infinite particle reservoirs at densities
$\rho_a$ and $\rho_b$.  As $\rho_a$ and $\rho_b$ are varied, the typical
macroscopic steady state density profile $\bar \rho(x)$,
$x\in[a,b]$, obtained in the limit $N=L(b-a)\to\infty$, 
exhibits shocks and phase transitions.  Here
we derive an exact asymptotic expression for the
probability of observing an arbitrary macroscopic profile $\rho(x)$:
$P_N(\{\rho(x)\})\sim\exp[-L{\cal F}_{[a,b]}(\{\rho(x)\});\rho_a,\rho_b]$, 
so  that ${\cal F}$ is the large deviation functional, 
a quantity similar
to the free energy of equilibrium systems. 
We find, as in the symmetric, purely diffusive case
$q=1$ (treated in an earlier  work), that $\cal F$ is in general a non-local
functional of $\rho(x)$.  Unlike the symmetric case, however, the asymmetric
case exhibits ranges of the parameters for which ${\cal F}(\{ \rho(x)\})$ is
not convex and others for which ${\cal F}(\{\rho(x)\})$ has discontinuities
in its second derivatives at $\rho(x) = \bar{\rho}(x)$; 
the fluctuations near $\bar{\rho}(x)$ are then non-Gaussian and cannot be
calculated from the large deviation function. 

\newpage

\section{Introduction\label{introduction}}

 Stationary nonequilibrium states (SNS) maintained by contact with infinite
thermal reservoirs at the system boundaries are objects of great
theoretical and practical interest \cite{dGM,HS,G,HS1,Sch,CH,DKS,Exp,ST}.
One is tempted to think that, at least when the gradients and fluxes
induced by the reservoirs are small, the full system behavior is just that
of a union of subsystems, each in local equilibrium, with spatially varying
particle and energy densities or equivalently local chemical potential and
temperature.  This is, however, not the entire story, as is clear when one
considers the paradigm of such systems, a fluid in contact with a thermal
reservoir at temperature $T_a$ at the top and one at temperature $T_b$ at
the bottom, the Rayleigh-B\'enard system \cite{CH}.  In this system there
are long range correlations not present in equilibrium systems, which have
been measured by neutron scattering experiments.  This system exhibits,
when $T_b-T_a$ exceeds some positive critical value, dynamic phase
transitions corresponding to the formation of different patterns of heat
and mass flow as the parameters are varied.  These are due to macroscopic
instabilities, caused by gravity, but are not derivable at present, despite
various attempts \cite{ST}, in terms of a microscopic theory of SNS such as
that provided by statistical mechanics in the case $T_a=T_b$, when the
system is an equilibrium one.

In this paper we study the SNS of a model system which, despite its
simplicity, has some phase transitions and exhibits many phenomena, such as
long range correlations and non-Gaussian fluctuations, which are very
different from those of systems in local thermal equilibrium.  For this
system the weights of the microscopic configurations are known and the typical
behavior has been deduced from them \cite{DEHP}.  Here we determine---again
from the microscopic weights---the probabilities of various atypical
macroscopic behaviors, that is, of large deviations. 

 The model we consider is the SNS of the open asymmetric simple exclusion
process (ASEP) \cite{Ligg1, ZS}: a lattice gas on a chain of N sites which
we index by $i$, $1 \leq i \leq N$.  At any given time $t$, each site is
either occupied by a single particle or is empty, and the system evolves
according to the following dynamics.  In the interior of the system ($2 \leq
i \leq N-1$), a particle attempts to jump to its right neighboring site with
rate 1 and to its left neighboring site with rate $q$ (with $0 \leq q < 1$). 
The jump is completed if the target site is empty, otherwise nothing happens. 
The boundary sites $i=1$ and $i=N$ are connected to particle reservoirs and
their dynamics is modified as follows: if site 1 is empty, it becomes
occupied at rate $\alpha$ by a particle from the left reservoir; if it is
occupied, the particle attempts to jump to site 2 (succeeding if this site is
empty) with rate 1.  Similarly, if site $N$ is occupied, the particle may
either jump out of the system (into the right reservoir) at rate $\beta$ or
to site $N-1$ at rate $q$. 

 More generally, one can also consider the ASEP with the above dynamical
rules supplemented by an output rate $\gamma$ at $i=1$ and an input rate
$\delta$ at $i=N$.  However, the calculations in the case of nonzero $\gamma$
or $\delta$ are more complicated, and for that reason we limit our analysis
in the present paper to the case $\gamma=\delta=0$. 

It is convenient to introduce the two parameters
 \begin{equation} \label{reservoir-asym}
  \rho_a = { \alpha \over 1 -q}\;, \ \ \ \rho_b = 1 - { \beta \over 1-q}\;;
  \end{equation}
 we require that $0\le\alpha,\beta\le1-q$, so that $0\le\rho_a,\rho_b\le1$. 
These parameters have a natural interpretation as reservoir densities.  In
particular, when $\rho_a=\rho_b$ the steady state of the system is a
Bernoulli measure at constant density $\rho_a$, that is, each site is
occupied independently with probability $\rho_a$; this may be seen for
example from the so-called ``matrix method'' (see
Section~\ref{matrix_method}).  In general, then, we interpret $\rho_a$ and
$\rho_b$ as the densities of particles in the left and right reservoirs,
respectively. 

The goal of the present work is to calculate the large $N$ behavior of
$P_N(\{\rho(x\})$, the probability of seeing a macroscopic profile $\rho(x)$
for $a < x< b$ in the steady state for a system of $N=L(b-a)$ sites.  This
probability $P_N(\{\rho(x\})$ can be thought of as the sum of the
probabilities of all microscopic configurations such that in each box of
 $L\,dx$ sites (with $dx \ll 1$ and $L\,dx \gg 1$),
the number of particles is close to $L \rho(x)\,dx$.
The ratio $\log( P_N(\{\rho(x\})) / L$ has a well defined limit for large
$L$,
 \begin{equation}
   \lim_{L\to \infty}{\log P_N(\{\rho(x\}) \over L}
  \equiv -{\cal F}_{[a,b]} (\{\rho(x)\};\rho_a,\rho_b)\;,
\label{calFdef}
  \end{equation}
 which depends on $b-a$, on the density profile $\rho(x)$, and on the
reservoir densities.  $\cal F$ is called the large deviation functional (LDF)
of the system. 

 When $\rho_a=\rho_b$, the steady state is a Bernoulli measure, as described
above.  This measure is just the equilibrium state of a system of particles,
noninteracting except for the hard-core exclusion, at chemical potential
$\log(\rho_a/(1-\rho_a))$.  For this system the LDF can be computed by
elementary means, and is given by \cite{Olla,Ellis,DLS2}
 \begin{equation}
 \label{Bernoulli}
 {\cal F}_{[a,b]} (\{\rho(x)\};\rho_a,\rho_a) = 
 \int_a^b \left[\rho(x) \log {\rho(x) \over \rho_a} 
    + (1 - \rho(x)) \log {1 - \rho(x) \over 1- \rho_a} \right] \,dx
 \end{equation}
 The present paper is devoted primarily to the derivation of the exact
expression of ${\cal F}$ in the case $\rho_a\ne\rho_b$. 

\subsection{Additivity and large deviations  in the ASEP}

 In our earlier work \cite{DLS,DLS2} on the large deviation functional for
the symmetric simple exclusion process, corresponding to equal jump rates
to the left and right, i.e., to $q=1$, we were able to use the matrix
method \cite{DEHP,Sandow,Sas,BECE} to calculate directly the probability of
a given macroscopic profile $\rho(x)$ by summing the probabilities of all
configurations corresponding to that profile.  For the asymmetric model,
however, such a direct calculation is {\it a priori} more complicated.  For
that reason, we follow here a different path, which has its origin in an
{\it a posteriori} observation made in \cite{DLS2}.  We noted there that
while the large deviation functional for the symmetric case is nonlocal, it
possesses a certain ``additivity'' property.  For the ASEP we first derive
an additivity property, similar to that of the symmetric model, and from
that obtain $\cal F$.  The derivations are given in Section
\ref{additivity}.

The addition formula involves a function ${\cal H}$ related to ${\cal F}$ by
 \begin{equation}
 \label{Hdef-asym}
 {\cal H}_{[a,b]}(\{\rho(x)\};\rho_a,\rho_b) 
  = {\cal F}_{[a,b]}(\{\rho(x)\};\rho_a,\rho_b) + (b-a) K(\rho_a,\rho_b)\;,
 \end{equation}
 where $K(\rho_a,\rho_b)$ does not depend on $\rho(x)$. 
 Since the dynamics in the bulk is driven from left to right (for $0\leq
q<1$), the roles played by the left and the right reservoirs
(\ref{reservoir-asym}) are not symmetric and the additivity relation
and the expressions for $K(\rho_a,\rho_b)$ in (\ref{Hdef-asym}) and for
${\cal F}_{[a,b]}(\{\rho(x)\};\rho_a,\rho_b)$ depend on whether $\rho_a >
\rho_b$ or $\rho_a < \rho_b$. 

\subsubsection{The case $\rho_a\ge\rho_b$}

 When $\rho_a\ge\rho_b$ the constant in (\ref{Hdef-asym}) is
 \begin{equation}
 K(\rho_a,\rho_b) =
 \sup_{ \rho_b \leq \rho \leq \rho_a}  \log [ \rho (1 - \rho) ],
\label{Kdef1}
 \end{equation}
 and the additivity relation, obtained in Section~\ref{additivity} below,
is that for any $c$ with $a<c<b$,
 \begin{eqnarray}
\label{additivity-asym1}
{\cal H}_{[a,b]} (\{\rho(x)\};\rho_a,\rho_b) \hskip-50pt && \\
 \nonumber &=&    \sup_{ \rho_b \leq \rho_c \leq \rho_a}
    \left[{\cal H}_{[a,c]} (\{\rho(x)\};\rho_a,\rho_c) +
     {\cal H}_{[c,b]} (\{\rho(x)\};\rho_c,\rho_b)\right] .
 \end{eqnarray}
  Equation (\ref{additivity-asym1}) expresses a relation between
the large deviation function of the whole system and those of two subsystems
connected at the break point to a reservoir at an appropriate density
$\rho_c$. 

 Once we have the additivity relation (\ref{additivity-asym1}), the
derivation of the large deviation functional is simple, and we give it here. 
We divide our system into $n$ parts of equal length and apply
(\ref{additivity-asym1}) $n$ times; this will introduce intermediate
reservoir densities
$\rho_a\equiv\rho_0\ge\rho_1\ge\cdots\ge\rho_n\equiv\rho_b$.  For very large
$n$, most of the intervals must have reservoir densities $\rho_{k-1},\rho_k$
at their boundaries which are nearly equal, and the LDF for these intervals
is approximately given by (\ref{Bernoulli}) (with $\rho_a$ there replaced by
$\rho_{k-1}\simeq\rho_k$).  On the other hand, the total length of the
intervals for which this is not true will approach $0$ for large $n$.  Now
taking the $n\to\infty$ limit and introducing a function $F(x)$ as the
interpolation of the values $\rho_0,\rho_1,\ldots,\rho_n$, we are lead
directly to a formula for the large deviation functional:
 \begin{eqnarray}
\label{result}
 {\cal F}_{[a,b]} (\{\rho(x)\};\rho_a,\rho_b) 
   = - (b-a) K(\rho_a,\rho_b)  \hskip-190pt &&  \\
\nonumber &+& \sup_{F(x)} \int_a^b dx\,
\rho(x) \log \left[ \rho(x) (1- F(x)) \right] +
(1 - \rho(x)) \log \left[ (1- \rho(x))  F(x) \right],
 \end{eqnarray}
 where the supremum is over all {\it monotone nonincreasing} functions $F(x)$
which for $a \leq x<y \leq b$ satisfy
 \begin{equation}
  \rho_a = F(a)  \ge F(x) \ge F(y) \ge F(b) = \rho_b.
\label{monotone}
 \end{equation}
 This supremum is achieved at a certain function $F_\rho(x)$.   
Note that without the constraints (\ref{monotone}) one would have
$F(x)=1-\rho(x)$.  The monotonicity requirement, however, makes the
determination of $F_\rho$ more subtle (see Section~\ref{Frho}) and the 
expression of ${\cal F}$ nonlocal.  This will be at the origin of most of its
interesting properties.  

\subsubsection{The case $\rho_a\le\rho_b$}

 When $\rho_a\le\rho_b$ the constant in (\ref{Hdef-asym}) is
 \begin{equation}  
K(\rho_a,\rho_b) =
\min \left[  \log  \rho_a (1 - \rho_a)
, \log \rho_b(1-\rho_b)  \right],
\label{Kdef2} 
 \end{equation}
 and the additivity relation, again derived in section~\ref{additivity}, is
 \begin{eqnarray}
 {\cal H}_{[a,b]} (\{\rho(x)\};\rho_a,\rho_b) 
 &=& \nonumber \\ 
 &&\hskip-80pt \min_{\rho_c=\rho_a,\rho_b}   
   \left[
    {\cal H}_{[a,c]} (\{\rho(x)\};\rho_a,\rho_c) +
    {\cal H}_{[c,b]} (\{\rho(x)\};\rho_c,\rho_b) \right].
\label{additivity-asym2}
\end{eqnarray}
 By an argument similar to that which led to (\ref{result}), we obtain 
the formula 
  \begin{eqnarray}
{\cal F}_{[a,b]}  (\{ \rho(x) \};\rho_a,\rho_b)) =   
- (b-a) K(\rho_a,\rho_b) + 
\hskip100pt
\label{result3} \\
   \inf_{a \leq y \leq b}
 \left\{ 
 \int_a^y dx\, \rho(x) \log \left[{ \rho(x)   (1 -\rho_a)} \right]
+ (1 - \rho(x)) \log \left[( 1- \rho(x))    \rho_a \right]
\right.
 \nonumber \\   
\left.
  +   \int_y^b dx\, \rho(x) \log \left[ \rho(x)  (1 -  \rho_b) \right]
+ (1 - \rho(x)) \log \left[ (1- \rho(x))    \rho_b \right]
 \right\} .
\nonumber
 \end{eqnarray}
 The fact that for each $\rho(x)$ one has to find in (\ref{result3}) the
infimum over $y$ makes, in this case too, the large deviation function
nonlocal. 

 It is interesting to note that the formulas (\ref{result}) and
(\ref{result3}) for the large deviation function do not depend on $q$.

\subsection{Outline of the paper}

 In Section~\ref{steady} we review briefly some known results on the open
ASEP, emphasizing the phase diagram.  In section~\ref{consequences} we give
a summary of our results which follow as consequences of the formulas
(\ref{result}) and (\ref{result3}) for the large deviation function.  In
Section~\ref{matrix_method} we recall the matrix method for carrying out exact
calculations in the ASEP and give several results obtained by this method
which are relevant to our considerations here.  Then in
Section~\ref{additivity} we give the derivation of the addition formulas
(\ref{additivity-asym1}) and (\ref{additivity-asym2}).  An unexpected
consequence of our results, discussed in Section~\ref{LD_Fluctuations}, is
that the correlations in the steady state are not related in any simple
manner to the large deviation function.  We also show in that section that
the fluctuations of the number of particles in any box of size $Lx$, with
$0<x<1$, are not Gaussian for certain ranges of the parameters
$\rho_a,\rho_b$.  Section~\ref{conclusion} gives some concluding remarks,
and certain more technical questions are discussed in appendices.

\section{The steady state of the ASEP with open boundary 
 conditions\label{steady}}

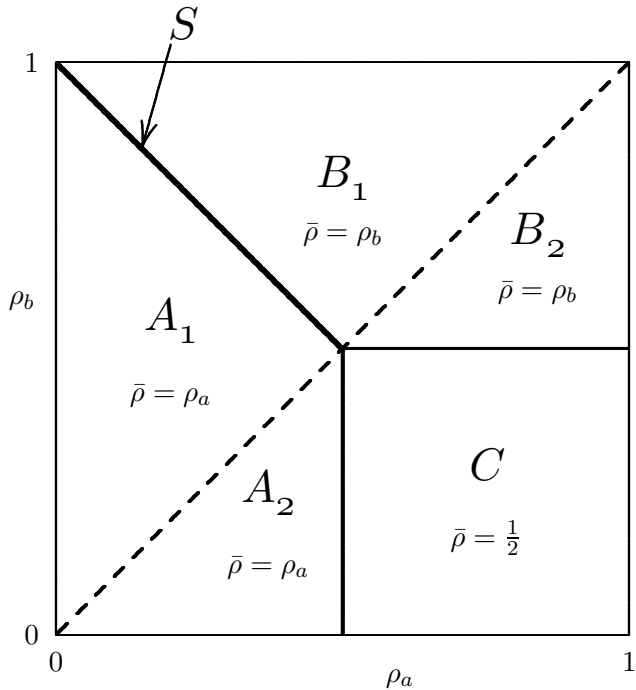
We describe here the phase diagram of the ASEP with open boundary
conditions, which has been obtained by various methods
\cite{krug,DEHP,Sas,BECE,SD}.  As indicated above, we consider here only
the case $\gamma=\delta=0$, $0\le\alpha,\beta\le1-q$.

\begin{figure}
\input phplane.tex
\end{figure}

The phase diagram is given in Figure~1, where we have chosen as parameters
the densities $\rho_a$ and $\rho_b$ (\ref{reservoir-asym}) of the two
reservoirs.  There are three phases: a low density phase $A$ with a constant
density $\bar{\rho}(x)= \rho_a$ in the bulk, a high density phase $B$ with a
density $\bar{\rho}(x)= \rho_b$, and a maximal current phase $C$ with a
density $\bar{\rho}(x)= 1/2$.  The current in each phase is given by
$J=(1-q)\bar\rho(1-\bar\rho)$.  The transition lines between these phases are
second order phase transitions where $\bar{\rho}(x)$ is a continuous function
of $\rho_a$ and $\rho_b$, except for the boundary $S$
 ($\rho_a=1 - \rho_b < 1/2$) between phase $A$ and $B$, where the transition
is first order: $\bar\rho(x)$ jumps from $\rho_a$ to $\rho_b$.  On the line
$S$ the typical configurations are shocks between phase $A$ with density
$\rho_a$ at the left of the shock and phase $B$ with density $\rho_b$ at the
right of the shock:
 \begin{equation}
  \rho_y(x)\equiv\rho_a \Theta(y-x) + \rho_b \Theta(x-y), \label{shock}
 \end{equation}
 with $\Theta$ the Heaviside function.  The position $y$ of the shock is
uniformly distributed along the system \cite{Ligg2,SA}, and as a result the
average profile is linear: $\langle\rhob(x)\rangle=\rho_a (1-x) + \rho_b x$.

This phase diagram can be understood easily in heuristic terms.  
First consider an infinite one dimensional lattice on which the initial
configuration is a Bernoulli distribution at density $\rho_a$ to the left of
the origin and $\rho_b$ to the right of the origin.  If $\rho_a < \rho_b$,
this initial condition produces a shock moving at velocity $(1- \rho_a -
\rho_b)(1-q)$; thus in the long time limit, the distribution near the origin
is Bernoulli, with density $\rho_a$ if $\rho_a + \rho_b < 1$ and
density $\rho_b$ if $\rho_a + \rho_b >1$.  On the other hand if $\rho_a >
\rho_b$, the profile becomes a rarefaction fan:
 $\rho(x,t)=\rho_a$ if $x\le x_a(t)$,
 $\rho(x,t)=\rho_a+(\rho_b-\rho_a)(x-x_a(t))/(x_b(t)-x_a(t))$
 if $x_a(t)<x<x_b(t)$, and $\rho(x,t)=\rho_b$ if $x_b(t)\le x$, with
 $x_\alpha(t) =(1-q)(1-2\rho_\alpha)t$, $\alpha=a,b$.
 If $1/2<\rho_b<\rho_a$ then the entire fan moves away to the left, if
$\rho_b<\rho_a<1/2$ then it moves away to the right, and if
$\rho_b<1/2<\rho_a$ then the origin remains in the fan for all time.  These
three cases give rise in the long time limit to Bernoulli distributions at
the origin, with densities $\rho_a$, $\rho_b$, and 1/2, respectively
(see e.g. \cite{Bramson} and references therein). 

If now we consider the finite system with left and right reservoirs at
densities $\rho_a$ and $\rho_b$, and start with a Bernoulli distribution at
density $\rho_a$ at the left of some point 
in the bulk far from the
boundaries, and $\rho_b$ at the right of this point, then the evolution will
be the same as in the infinite system until the shock or the fan reaches the
boundary, leading, for the asymptotic density in the bulk, to what is given
in the phase diagram.  This idea is at the basis of what has been done
recently by Popkov and Sch\"utz \cite{PS,PSbis} to predict boundary induced
phase diagrams in more general cases.

 The dashed line $\rho_a=\rho_b$ ($\alpha+ \beta = 1-q$) in Figure~1
separates what we will call the {\it shock region} $\rho_a<\rho_b$ and the
{\it fan region} $\rho_a>\rho_b$.  On this line the measure reduces to a
Bernoulli measure at density $\rho_a$ (see Section~\ref{matrix_method}) and
the large deviation function is given by (\ref{Bernoulli}).  The line plays
no role in the phase diagram for the typical profile $\bar\rho$ but separates
phases $A$ and $B$ into two subphases, $A_1,A_2$ and $B_1,B_2$, which, as we
have seen in Section~\ref{introduction}, can be distinguished by the
different expressions (\ref{result}), (\ref{result3}) for the large deviation
function in these regions.

\section{Consequences of the large deviation formula for the
ASEP\label{consequences}}

In this section we describe some consequences of formulas (\ref{result}) and
(\ref{result3}) for the large deviation function.  It is convenient to write
(\ref{Kdef1}) and (\ref{Kdef2}) in the unified form
 \begin{equation}
 \label{simpleK}
  K(\rho_a,\rho_b)= \log\bar\rho(1-\bar\rho),
 \end{equation}
 where $\bar\rho=\bar\rho(x)$ depends on $\rho_a$ and $\rho_b$ and is
obtained from the phase diagram, Figure 1.  Note that $\bar\rho$ is in fact
independent of $x$ except on the line $S$; there $\bar\rho$ can be any shock
profile $\rho_y(x)$ (\ref{shock}), but the value of $K(\rho_a,\rho_b)$ is
independent of the shock position $y$ since $\rho_a+\rho_b=1$ on $S$. 
  We also introduce the notation
 \begin{equation}
  \label{hdef}
  h(r,f;\bar\rho)
   = r\log{r\over f}+(1-r)\log{1-r\over1-f}
   +\log{f(1-f)\over\bar\rho(1-\bar\rho)},
 \end{equation}
 so that (\ref{result}) and (\ref{result3}) become respectively 
 \begin{eqnarray}
 \label{result_h}
 {\cal F}_{[a,b]} (\{\rho(x)\};\rho_a,\rho_b) &&\nonumber\\
   && \hskip-85pt =\; \sup_{F(x)} \int_a^b dx\,h(\rho(x),F(x);\bar\rho)
   = \int_a^b dx\,h(\rho(x),F_\rho(x);\bar\rho),
 \end{eqnarray}
 for $\rho_a>\rho_b$, and 
 \begin{eqnarray}
{\cal F}_{[a,b]}  (\{ \rho(x) \};\rho_a,\rho_b)&&\nonumber\\
 \label{result3_h}
  &&\hskip-85pt =\; \inf_{a \leq y \leq b}
 \left\{ \int_a^y dx\,\,h(\rho(x),\rho_a;\bar\rho)
  +\int_y^b dx\,\,h(\rho(x),\rho_b;\bar\rho)\right\},
 \end{eqnarray}
 for $\rho_a<\rho_b$, where again $\bar\rho$ is determined from $\rho_a,\rho_b$ through the phase
diagram.  Note that $h(r,f;\bar\rho)$ is strictly convex in $r$ for fixed
$f,\bar\rho$, with a minimum at $r=f$.

 \subsection{Construction of the function $F_\rho$\label{Frho}}

There is a rather simple way of constructing the optimizing function 
$F_\rho(x)$ in (\ref{result_h}).  Let $G_\rho(x)$ be defined for
$a\le x\le b$ by
 \begin{equation}
 G_\rho(x) 
  =  \mbox{Concave Envelope}\left\{ \int_a^ x (1- \rho(y)) dy \right\};
\label{CE}
 \end{equation}
 then $F_\rho$ is obtained by cutting off $G'_\rho(x)$ at $\rho_a$ and
$\rho_b$: 
 \begin{equation}\label{COCE}
 F_\rho(x) =  \left\{
    \begin{array}{ll}
      \displaystyle {\rho_a},& \mbox{if $G'(x) \geq \rho_a .$} \\
      \displaystyle {{ G'(x) } }, & \mbox{if $\rho_b\leq G'(x)\leq\rho_a$,}\\
      \displaystyle {\rho_b},& \mbox{if $G'(x) \leq \rho_b,$} 
    \end{array}
\right.
 \end{equation}
This construction is verified in Appendix~\ref{con_env}.

 Suppose, for example, that $\rho(x)$ is a constant profile: $\rho(x)=r$. 
Then the expression in brackets in (\ref{CE}) is concave and hence equal to
$G_\rho(x)$, so that $G'_\rho(x)=1-r$ and $F_\rho$
is constant: $F_\rho=\rho_a$ if $r<1-\rho_a$, $F_\rho=1-r$ if
 $1-\rho_a\le r\le1-\rho_b$, and $F_\rho=\rho_b$ if $1-\rho_b<r$.  In
particular, taking $r=\bar\rho$ we find that
 \begin{equation}
  F_{\bar\rho}=\bar\rho.\label{Frhobar}
 \end{equation}
 For (i)~in region
$A_2$, where $1/2>\rho_a>\rho_b$, $r=\rho_a<1-\rho_a$ and so
$F_\rho=\rho_a=\bar\rho$; (ii)~in region
$C$, where $\rho_a>1/2>\rho_b$, $r=1/2$ and so
$F_\rho=1-r=1/2=\bar\rho$; (iii)~~in region
$B_2$, where $\rho_a>\rho_b>1/2$, $r=\rho_b>1-\rho_b$ and so
$F_\rho=\rho_b=\bar\rho$.

 \subsection{The most likely profile\label{MLP}}

We will show in this section that the most likely profile is always given by
$\rho(x)=\bar\rho$; specifically, that 
${\cal F}_{[a,b]}(\{\rho(x)\};\rho_a,\rho_b)\ge0$ for all $\rho(x)$, that
${\cal F}_{[a,b]}(\{\bar\rho\};\rho_a,\rho_b)=0$, and that 
${\cal F}_{[a,b]}(\{\rho(x)\};\rho_a,\rho_b)>0$ if $\rho(x)\ne\bar\rho$.
This is of course expected and in fact represents an alternate way to obtain
the phase diagram of Figure~1. 

  To obtain the most likely profile in the fan region $\rho_a\ge\rho_b$, we
note that (\ref{result_h}), (\ref{Frhobar}), and the
convexity of $h(r,f;\bar\rho)$ in $r$ imply that 
 \begin{eqnarray}
 {\cal F}_{[a,b]}(\{\rho(x)\};\rho_a,\rho_b)
    &=&    \int_a^b dx\,h(\rho(x),F_\rho(x);\bar\rho)\nonumber\\
    &\ge&   \int_a^b dx\,h(\rho(x),F_{\bar\rho};\bar\rho)\nonumber\\
    &\ge&   \int_a^b dx\,h(F_{\bar\rho},F_{\bar\rho};\bar\rho)
   = 0. \label{minfan}
 \end{eqnarray}
  Moreover, if $\rho(x)=\bar\rho$ then equality holds throughout
(\ref{minfan}), so that the minimum value of $\cal F$ is zero and $\bar\rho$
is a minimizer; otherwise the second inequality is strict, from which it
follows that this minimizer is unique. 

  In the shock region $\rho_a<\rho_b$, we observe that for fixed $y$, $a\le
y\le b$, the right side of (\ref{result3_h}) is minimized by the unique
choice $\rho(x)=\rho_a$ for $a\le x<y$, $\rho(x)=\rho_b$ for $y<x\le b$. 
Minimizing over $y$ then implies that, except on the first order line $S$,
the optimal profile is again constant with value $\bar\rho$, and is unique;
again the corresponding minimum value of $\cal F$ is zero.  On $S$ all values
of $y$ give the value zero for $\cal F$, so that the shock profiles
$\rho_y(x)$ of (\ref{shock}) form a one parameter family of minimizing
profiles.  Note that a knowledge of the large deviation functional is not
sufficient to determine the distribution of the shock position $y$ mentioned
in Section~\ref{steady}.

 \subsection{Convexity\label{convexity}}

 In the fan region $\rho_a \geq \rho_b$, the LDF ${\cal
F}_{[a,b]}(\{\rho(x)\};\rho_a,\rho_b)$ is a strictly convex functional of
$\rho(x)$, since by (\ref{result_h}) it is the maximum (over the functions
$F(x)$) of strictly convex functionals of $\rho(x)$.  This is also true
in the symmetric case \cite{DLS,DLS2} and in
equilibrium systems not at a phase transition.
In the shock region $\rho_a < \rho_b$, on the contrary,
$\cal F$ is not convex.  This is most easily verified on the line $S$, since
it follows from Section~\ref{MLP} that on $S$ a superposition of minimizing
profiles (\ref{shock}), $\rho(x)=\lambda\rho_y(x)+(1-\lambda)\rho_z(x)$, $y\ne z$,
satisfies ${\cal F}(\{\rho(x)\})>0$ for $0<\lambda<1$.  But we will also see
in Section~\ref{flat} below that for every $\rho_a,\rho_b$ there is a
constant profile $\rho(x)=r^*$ near which $\cal F$ is not convex. 

 \subsection{Suppression and enhancement of large deviations\label{ESLD}}

 The LDF in the fan region $\rho_a>\rho_b$ has similarities besides
convexity to the LDF in the symmetric case.  In particular it is easy to
see from (\ref{result}) that
 \begin{equation}
 \label{suppress}
{\cal F}_{[a,b]}(\{\rho(x)\};\rho_a,\rho_b) 
     \geq {\cal F}_{[a,b]}^{\rm eq}(\{\rho(x)\};\bar\rho),
  \qquad \hbox{if $\rho_a\ge\rho_b$},
 \end{equation}
 where we define
 \begin{equation}
 \label{equilibrium}
 {\cal F}_{[a,b]}^{\rm eq}(\{\rho(x)\};\bar\rho)
  = \int_a^b \left[\rho(x) \log {\rho(x) \over \bar\rho} 
    + (1 - \rho(x)) \log {1 - \rho(x) \over 1- \bar\rho} \right] \,dx\;;
 \end{equation}
 this is the LDF for an equilibrium system at density $\bar\rho$ 
(see (\ref{Bernoulli}))  But in the shock region this
inequality is reversed:
 \begin{equation}
 \label{enhance}
{\cal F}_{[a,b]}(\{\rho(x)\};\rho_a,\rho_b) 
     \leq {\cal F}_{[a,b]}^{\rm eq}(\{\rho(x)\};\bar\rho),
  \qquad \hbox{if $\rho_a\le\rho_b$},
 \end{equation}
 as can be derived from (\ref{simpleK}) and (\ref{result3_h}), since in
region $B_1$ ($A_1$), taking $y=a$ ($y=b$) on the right side of
(\ref{result3_h}) gives ${\cal F}_{[a,b]}^{\rm eq}(\{\rho(x)\};\bar\rho)$. 
On the line $S$, where $\bar\rho=\rho_y(x)$ (see (\ref{shock})) for some $y$,
(\ref{enhance}) holds for all values of $y$.  Physically, (\ref{suppress})
and (\ref{enhance}) mean that the probability of a macroscopic deviation from
the typical density profile is reduced in the fan region, and increased in
the shock region, compared with the probability of the same deviation in an
equilibrium system with the same typical profile. 

 \subsection{Flat density profiles\label{flat}}

 One can readily compute the LDF for constant profiles $\rho(x)=r$.  In the
fan region $\rho_a > \rho_b$ one finds, as discussed in Section~\ref{Frho},
that $F_\rho$ is also constant, with $F_\rho=\rho_a$ if $r<1-\rho_a$,
$F_\rho=1-r$ if $1-\rho_a\le r\le1-\rho_b$, and $F_\rho=\rho_b$ if
$1-\rho_b<r$.  There are correspondingly three different expressions for the
LDF $\widehat{\cal F}(r)\equiv{\cal F}_{[a,b]}(r;\rho_a,\rho_b)$:
 \begin{equation}
  \widehat{\cal F}(r)
 = \left\{
    \begin{array}{ll}
      (b-a)h(r,\rho_a;\bar\rho),& \mbox{if $r<1-\rho_a$,} \\
      (b-a)h(r,1-r;\bar\rho), & \mbox{if $1-\rho_a\le r\le1-\rho_b$,} \\
      (b-a)h(r,\rho_b;\bar\rho),& \mbox{if $1-\rho_b<r$.} 
    \end{array}
\right.
 \label{flat1}
 \end{equation}

 If $\rho_a>1/2>\rho_b$, so that $\bar\rho=1/2$, (phase C of Figure~1) and
$1-\rho_a\le r\le1-\rho_b$, then from (\ref{flat1}),
 \begin{equation}\label{factor2}
  \widehat{\cal F}(r)=2{\cal F}_{[a,b]}(r;1/2,1/2)
  = 2{\cal F}_{[a,b]}^{\rm eq}(r;1/2),
 \end{equation}
  where again (see (\ref{equilibrium})) ${\cal F}_{[a,b]}^{\rm eq}(r;1/2)$ is
the large deviation function for observing the uniform density $r$ in a
Bernoulli measure with density $1/2$.  In particular, for $\rho_a=1$,
$\rho_b=0$, the probability of observing all sites empty, $r=0$, or all sites
occupied, $r=1$, is $2^{-N}$ for the Bernoulli measure, where $N=L(b-a)$ is
the number of sites, so that from (\ref{factor2}) the corresponding
probability for the ASEP is given to leading order by $4^{-N}$. 

 In the shock region $\rho_a < \rho_b$, the minimizing  $y$ in
(\ref{result3}) is $y=b$ if $r<r^*$ and $y=a$ if $r>r^*$, where
 \begin{equation}
r^*= {\log{\rho_b \over \rho_a} \over \log \left( {1-\rho_a \over 1 -\rho_b} {\rho_b \over \rho_a} \right) }
\label{rstar}
 \end{equation}
so that
 \begin{equation}
  \widehat{\cal F}(r)
 = \left\{
    \begin{array}{ll}
      (b-a)h(r,\rho_a;\bar\rho),& \mbox{if $r\le r^*$,} \\
      (b-a)h(r,\rho_b;\bar\rho),& \mbox{if $r\ge r^*$.}
    \end{array}
\right.
 \label{flat2}
 \end{equation}
 At $r=r^*$ the derivative is discontinuous, with
 \begin{equation}
{d \widehat{\cal F}\over dr}\bigg|_{r\to r^*-0}
 > {d \widehat{\cal F}\over dr}\bigg|_{r\to r^*+0}.
 \end{equation}
 Thus $\widehat{\cal F}(r)$ is not convex near $r=r^*$,
and hence ${\cal F}_{[a,b]}(\{\rho(x)\};\rho_a,\rho_b)$) is not convex 
in a neighborhood of $\rho(x)=r^*$.

 \subsection{Distribution of the total number of particles\label{number}}

 The probability $P_L(M)$ that there are a total of $M=rN$ particles 
in the system can be obtained from the LDF as
 \begin{equation}
-{1\over L}P_L(rN)\simeq \widetilde{\cal F}(r)
   \equiv{\cal F}_{[a,b]}(\{\rho(x)\};\rho_a,\rho_b),
 \end{equation}
 where $\rho(x)$ is the most likely profile under the constraint
 \begin{equation}\label{constraint0}
\int_a^b \rho(x)\,dx = r(b-a).
 \end{equation}
 It will be shown in Appendix~\ref{opt_constraint} that $\rho(x)$ is the
constant profile $\rho(x)=r$, and correspondingly $\widetilde{\cal
F}(r)=\widehat{\cal F}(r)$,  except in that portion of the shock region in
which $r$ satisfies $1-\rho_b < r < 1 - \rho_a$, where 
 \begin{equation}
  \label{constrained-shock}
\rho(x) =  \cases{ 1- \rho_b, & \mbox{if $x < y_r$ },\cr
                     1-   \rho_a, & \mbox{if $x > y_r$,  }} 
 \end{equation}
 with 
 \begin{equation}
 \label{y_crit}
 y_r={b (1- \rho_a - r) - a ( 1 - \rho_b -r) \over \rho_b - \rho_a}.
 \end{equation}
 Then from (\ref{result3_h}) it follows that for these values of $r$,
 \begin{equation}
\widetilde{\cal F}(r)
   = (b-a) [ r \log (1-\rho_a) (1- \rho_b) + (1-r) \log \rho_a \rho_b 
      - K(\rho_a,\rho_b)].
\label{Ftilde}
 \end{equation}

 {\bf Remarks:} (a) Equation (\ref{constrained-shock}) is easy to interpret on
the first order line $S$ in the phase plane, where $\rho_a+\rho_b=1$: it
corresponds there to a typical shock configuration $\rho_{y_r}(x)$, with
$y_r$ determined by (\ref{constraint0}). 

(b) Although, as observed in Section~\ref{flat}, $\widehat{\cal F}$ is not
convex (at least in the shock region), $\widetilde{\cal F} $ is always
convex, and is in fact the convex envelope of $\widehat{\cal F}$. 

(c) We will discuss in Section~\ref{nongaussian} how (\ref{Ftilde}) may be
derived by a direct calculation from the matrix method.

 \subsection{Small fluctuations\label{SF}}

It is natural to ask about the connection between the LDF, which gives the
probabilities of macroscopic deviations from the typical density profile, and
the distribution of small fluctuations, i.e., those of order $1/\sqrt N$. 
(In what follows we will refer to these simply as ``fluctuations.'') In the
symmetric case discussed in \cite{DLS2}, as in an equilibrium system not at a
phase transition (in any dimension, with $N$ being the number of sites in the
system), the distribution of fluctuations can be obtained from ${\cal F}$ as
a limit.  More precisely if we write $\rho(x) = \bar \rho(x) + {1 \over \sqrt
L} u(x)$ and then expand $\cal F$ to second order (the first order term being
zero) we get a Gaussian distribution for $u(x)$ with covariance $C(x,x')$,
where $C^{-1}(x,x') = \delta^2 {\cal F}/\delta \rho(x) \delta \rho(x')$
evaluated at $\rho = \bar \rho$.  This covariance is the suitably scaled
microscopic truncated pair correlation \cite{DLS2}.  

For the asymmetric case discussed in this paper, however, the distribution of
small fluctuations need no longer be given by the LDF, as we discuss in
Section~\ref{Discontinuous_LD}.  In fact we show there that $\delta^2 {\cal
F}/\delta \rho (x) \delta \rho(x')$ is discontinuous at $\bar \rho = 1/2$ in
the interior of region $C$ of the phase diagram, i.e., where
$\rho_a>1/2>\rho_b$.  Furthermore, the fluctuations in this region are no
longer Gaussian; in Section~\ref{nongaussian} we show this by computing
explicitly the non-Gaussian distribution of the fluctuations of the number of
particles in a box of size $Ly$, with $0<y<1$. 

\section{The matrix method for the ASEP\label{matrix_method}}

The steady state properties of the ASEP with open boundaries can be
calculated exactly in various ways \cite{DDM,DE,SD}; here we describe the
so-called matrix method \cite{DEHP,Sandow,Sas,BECE}, which we will use in
the derivation of the additivity relations in Section~\ref{additivity}.  Let us
consider two operators denoted by $D$ and $E$, a left vector $\langle W|$ and
a right vector $|V\rangle$, which satisfy the following algebraic rules:
 \begin{eqnarray}
     DE - qED &=&  D + E  \label{DE}\;,\label{alg1}\\
  \beta D |V\rangle &=&  |V\rangle\;,\label{alg2}\\
   \langle W|\alpha E  &=&  \langle W|\;\label{alg3}.
  \end{eqnarray}
 Any matrix element of the form
$ \langle W | Y_1 Y_2 \cdots Y_k| V\rangle  /  \langle W |  V\rangle$, 
 where $Y_i$ denotes $D$ or $E$, can be calculated from these rules (without
the need of writing down an explicit representation).  Thus from
(\ref{alg2},\ref{alg3}), one has
 $\langle W |D | V \rangle / \langle W | V \rangle = 1/\beta$ and
 $\langle W |E | V \rangle / \langle W | V \rangle = 1/\alpha$,  and 
the matrix element of any product of $n$ matrices can
be expressed through (\ref{alg1})--(\ref{alg3}) in terms of sums of elements of
shorter products.

For the open ASEP as described in Section~\ref{introduction}, the probability
$P(\{\tau_i\})$ of the microscopic configuration $\{\tau_i\}$ in the steady
state can be written as \cite{DEHP}
 \begin{equation}
 P(\{\tau_i\}) 
  = { \langle W | \prod_{i=1}^N ( D \tau_i + E(1-\tau_i) ) | V \rangle 
      \over Z_N}  ,
\label{matrix}
 \end{equation}
where $\tau_i=1$ or $0$ indicates whether site $i$ is occupied  or empty 
and the normalization factor $Z_N$ is given by
 \begin{equation}
Z_N = \langle W | (D+E)^N | V \rangle .
\label{normalisation}
 \end{equation}

 On the dashed line $\rho_a=\rho_b$ of Figure 1, there exists a one
dimensional representation of the matrix algebra; it then follows
immediately from (\ref{matrix}) that the invariant measure is Bernoulli on
this line.  There are other lines in the $\rho_a,\rho_b$ plane where there
exist finite dimensional representations \cite{ER,MS} of the algebra
(\ref{alg1})--(\ref{alg3}), but the large deviation function has no
particular or remarkable expression along these lines, and so these cases
will be treated together with the general case.

 From the algebra (\ref{alg1})--(\ref{alg3}) all (equal time) steady state
properties can (in principle) be calculated.  For example the average
occupation $\langle \tau_i \rangle $ of site $i$ is given by
 \begin{equation}
\langle \tau_i \rangle  
  = {\langle W | (D+E)^{i-1} D (D+E)^{N-i} | V \rangle \over Z_N},
\label{taui}
 \end{equation}
and the two point function is, for $i < j$,
 \begin{equation}
\langle \tau_i \tau_j\rangle  
 = {\langle W | (D+E)^{i-1} D (D+E)^{j-i-1} D (D+E)^{N-j} | V \rangle 
   \over Z_N}.
\label{tauitauj}
 \end{equation}
 The average steady state current $J_N$ is the same across any bond.  It has
the form
 \begin{eqnarray}
   J_N  &=& \langle \tau_i (1 - \tau_{i+1} ) 
     - q (1 - \tau_i) \tau_{i+1}  \rangle \hskip100pt \nonumber \\
   &=&  {\langle W | (D+E)^{i-1} (DE - q ED)  (D+E)^{N-i-1} | V \rangle 
     \over Z_N} = {Z_{N-1} \over Z_N}.
\label{current}
 \end{eqnarray}
 The probability $Q_{N_1,..N_k}(M_1,..M_k)$ that there are exactly $M_1$
particles on the first $N_1$ sites, $M_2$ particles on the next $N_2$ sites,
\dots, $M_k$ particles in the rightmost $N_k$ sites is given by
 \begin{equation}
 Q_{N_1,\ldots,N_k}(M_1,\ldots,M_k)=
 {\langle W | Y_1\cdots Y_k | V \rangle \over Z_N} ,
\label{Q}
 \end{equation}
with 
 \begin{equation}
Y_p = {1 \over 2 i \pi} \int_0^{2 \pi} d \theta
 \ e^{- i \theta M_p}  \ 
 (D e^{i \theta} + E)^{N_p} . 
 \end{equation}
  Clearly by making the number $k$ of boxes, and their sizes $N_p$, large
enough, we can approximate any density profile $\rho(x)$ via
 $\rho(x)= M_p/N_p$ for $x= p/k$.  Equation (\ref{Q}) then gives the
probability of the profile $\rho(x)$, and one may attempt to obtain an
asymptotic form via a saddle point analysis.  This approach, which is
conceptually straightforward and was followed in the symmetric case
\cite{DLS2}, turns out to be more difficult to implement in the asymmetric
case, and this is why we here follow the additivity approach explained in
Section~\ref{additivity}. 

 Various more explicit expressions may be extracted from (\ref{taui}) and
(\ref{tauitauj}); for example, in the totally asymmetric case ($q=0$) 
the average profile
 $\langle \tau_i \rangle$ boundaries was computed for all $i$ in \cite{DEHP}. 
Moreover, for $q=0$ and at the special point $\alpha=\beta=1$ ($\rho_a=1$ and
$\rho_b=0$) in the maximal current phase $C$, finite-size corrections have
been computed \cite{DE} for the mean density profile $\langle \tau_i \rangle$
and for the two point function $\langle \tau_i \tau_j \rangle - \langle
\tau_i \rangle \langle \tau_j \rangle$.  In particular it was found in
\cite{DE} that at a point $i = Nx$ (with $1 \leq i \leq N$) of a system of
$N$ sites
 \begin{equation}
\langle  \tau_i \rangle = {1 \over 2} + {1 \over 2 \sqrt{\pi}} {1 \over N^{1/2}} {1-2x \over \sqrt{x(1-x)}} + O(N^{-3/2}) 
 \end{equation} 
which for a system of $N= L(b-a)$ sites  becomes for $i= L(x-a)$
 \begin{equation}
 \langle  \tau_i \rangle 
  = {1\over 2} + {1\over 2 \sqrt{\pi}}{1 \over L^{1/2}}
         {b+a - 2 x  \over \sqrt{(b-a)(b-x)(x-a)}} + O(L^{-3/2}) 
\label{tauiasymp}
 \end{equation}

 Also, at this point ($ \alpha = \beta =1$) of the phase diagram it was shown
\cite{DE} that for large $N$, the variance of the total number $M$ of
particles is given by
 \begin{equation}
\label{varianceM}
\langle M^2 \rangle - \langle M \rangle^2 \simeq {N \over 8}
 \end{equation}
while a Bernoulli distribution at density $1/2$ would give twice this
variance (see the discussion in Section~\ref{flat}).

\section{Derivation of the additivity 
 formulae \label{additivity}}

In this section we obtain the additivity formulae (\ref{additivity-asym1})
and (\ref{additivity-asym2}) for the large deviation functional in the ASEP. 
We begin by deriving two formulae valid for arbitrary system size $N$.  The
first of these, (\ref{main-claimbis}), is obtained directly from the matrix
formalism of Section~\ref{matrix_method} and is valid for a range of
parameters corresponding to the fan region $\rho_a > \rho_b$.  By analytic
continuation of (\ref{main-claimbis}) we then obtain a new formula
(\ref{main-claim4}), which is valid for all parameter values.  Finally, we
analyze the large $N$ behavior of (\ref{main-claim4}) to obtain
(\ref{additivity-asym1}) and (\ref{additivity-asym2}). 

For $q=0$, the additivity formula (\ref{main-claimbis}) takes a much simpler
form, which is given in \cite{DLSasep}.  We do not reproduce this formula in
the current paper because the treatment here of the general case $0\le q < 1$
requires slightly different notations.

\subsection{Preliminaries\label{prelims}}

Let $D$ and $E$ satisfy (\ref{alg1})--(\ref{alg3}) with 
$q<1$. We define the operators $d$ and $e$ by
 \begin{equation}
D = {1 \over 1-q} (1+  d)\;,   \qquad  E = {1 \over 1-q} (1 + e)\;;
\label{dedef}
 \end{equation}
 from (\ref{alg1}) these operators satisfy
 \begin{equation}
d \ e -q \  e \ d =   1-q . 
\label{alg1bis}
 \end{equation}
We also define eigenvectors $|z\rangle $ and $\langle z|$ of $d$ and
$e$, for arbitrary complex $z$, by
 \begin{eqnarray}
d \ |z \rangle = z \  |z \rangle \;,
\label{vdef}
\\
\langle z | \ e = {1\over z} \ \langle z | \;.
\label{wdef}
 \end{eqnarray}
 If $X$ is a polynomial in the operators $D$ and $E$
(or equivalently $d$ and $e$), then the matrix element
$\langle z_0|X|z_1\rangle/\langle z_0|z_1\rangle$
 is a polynomial in $1/z_0$ and $z_1$, with positive
coefficients whenever the coefficients in $X$ are positive; this is easily
seen since then $X$ is a polynomial in $d$ and $e$ with positive
coefficients, and using (\ref{alg1bis}) one can push all the $d$'s to the
right and all the $e$'s to the left, maintaining this positivity.

 Finally, we define the function $\varphi(z)$ by
 \begin{equation}
\varphi(z) =  \sum_{n=0}^\infty  {1 \over c_n }\ z^n\;,
\label{phidef}
 \end{equation}
where the $c_n$ are constructed from the recursion
 \begin{equation}
c_0 = 1\;, \qquad  c_{n} = (1 - q^n) \  c_{n-1}.
\label{cndef}
 \end{equation}
One can check easily from (\ref{phidef})  
that  
 \begin{equation}
 \label{fctl_eqn}
\varphi(z)-\varphi( qz)= z \varphi(z)\;,
 \end{equation}
 so that
 \begin{equation}
 \varphi(z) = \prod_{n=0}^\infty {1 \over 1 - q^n z} \;.
\label{phiexp}
 \end{equation}

\subsection{Exact additivity formula for $|z_0| > |z_1|$\label{add_fan}}

The additivity formula which we prove in this section is that if $X_0$ and
$X_1$ are arbitrary polynomials in the operators $D$ and $E$, then for
 \begin{equation}
|z_0| > |z_1|
\label{z0gz1}
 \end{equation}
 one has 
 \begin{eqnarray}
 \label{main-claimbis}
 {\langle z_0|X_0X_1|z_1\rangle\over\langle z_0|z_1\rangle}
    \varphi \left( z_1 \over z_0 \right)&&\\
 &&\hskip-110pt\;=  
  {1\over\varphi(q)}
   \sum_{n=0}^\infty  {q^n \over c_n}
   \oint {dz \over 2 \pi i z}
    {\langle z_0|X_0|q^nz\rangle\over\langle z_0|q^nz\rangle}
       \varphi \left( q^n z \over z_0 \right)
 {\langle z|X_1|z_1\rangle\over\langle z|z_1\rangle}
   \varphi \left( z_1 \over z \right)\;,
 \nonumber
 \end{eqnarray}
where the contour of integration is a circle $|z|=R$ with
 \begin{equation}
| z_1 | < R  < |z_0|\;.
\label{conditionbis}
 \end{equation}
{\bf Proof:} Using (\ref{alg1bis}), any polynomial $X$ in $D$ and $E$
can be written in the form
 \begin{equation}
X= \sum_{p,p'} A_{p,p'} \ e^{p'}\  d^p\;.
 \end{equation}
 As (\ref{main-claimbis}) is linear in $X_0$ and $X_1$, it is sufficient to
prove it for $X_0 = e^{p_0'} d^{p_0}$ and $X_1 = e^{p_1} d^{p_1'}$, and as
$\langle z_0|$ and $ |z_1\rangle$ are
eigenvectors of $e$ and $d$, one can immediately simplify the problem and
limit the discussion to the case
 \begin{equation}
X_0 = d^{p_0}\;, \qquad  X_1= e^{p_1}\;.
\label{choice}
 \end{equation}
 For this choice of $X_0$ and $X_1$, the right hand side of
(\ref{main-claimbis}) becomes
 \begin{equation}
   {1 \over \varphi(q)} \sum_{n=0}^\infty \sum_{m_0=0}^\infty
     \sum_{m_1=0}^\infty {q^{n(1+m_0+p_0)} \over c_n \  c_{m_0} \  c_{m_1}} 
   \  {z_1^{m_1} \over z_0^{m_0}} \  \delta_{m_0+p_0, m_1+p_1}\;,
 \end{equation}
 once $ \varphi(q^n z/z_0)$ and $\varphi ( z_1/z )$ have been replaced by
their power series (\ref{phidef}) and the integration over $z$ has been
carried out.  One can now use (\ref{phidef}) again to evaluate the sum over
$n$, and then simplify the result using the identity
$\varphi(q^{1+m})=c_m\varphi(q)$, which follows from (\ref{fctl_eqn}) by an
inductive argument, to obtain
 \begin{eqnarray}
 \mbox{r.h.s.  of (\ref{main-claimbis})} 
  &=& \sum_{m_0=0}^\infty \sum_{m_1=0}^\infty
   {\varphi(q^{1+m_0+p_0}) \over \varphi(q)} \ {1 \over
c_{m_0}\ c_{m_1}}\ {z_1^{m_1} \over z_0^{m_0}}\ \delta_{m_0+p_0, m_1+p_1} ,
 \nonumber\\
 \label{rhs}
  &=& \sum_{m_0=0}^\infty \sum_{m_1=0}^\infty 
   {c_{m_0+p_0} \over  c_{m_0} c_{m_1}} \ 
    {z_1^{m_1} \over z_0^{m_0}} \  \delta_{m_0+p_0, m_1+p_1}\;.
 \end{eqnarray}
 So to prove (\ref{main-claimbis}), we just need to show that the left hand
side of (\ref{main-claimbis}) coincides with (\ref{rhs}). 

We argue by induction on $p_0$.  If $p_0=0$, it is easy to see that the left
hand side of (\ref{main-claimbis}) is given by
$z_0^{-p_1}\varphi({z_1/z_0})$.  Evaluating (\ref{rhs}) in this case leads
to the same expression. 

Let us assume, then, that (\ref{main-claimbis}) is valid for all
 $p_0\leq P$.  We want to prove that it remains true for $p_0=P+1$.  To do so
we observe the following consequence of (\ref{alg1bis}):
 \begin{equation} d^{P+1} \ e^{p_1} = (1 - q^{p_1})  d^P e^{p_1-1} + q^{p_1}
d^P e^{p_1} d \;.
\label{identity}
 \end{equation}
 Using (\ref{identity}), the left hand side of (\ref{main-claimbis}) for
$p_0=P+1$, i.e., for $X_0=d^{P+1}$ and $X_1=e^{p_1}$, becomes
 \begin{eqnarray}
  (1-q^{p_1})
  {\langle z_0|d^{P}e^{p_1 -1}|z_1\rangle\over\langle z_0|z_1\rangle}
   \varphi\left(z_1\over z_0\right)
   +  q^{p_1} 
  {\langle z_0|d^{P}e^{p_1}d|z_1\rangle\over\langle z_0|z_1\rangle}
  \varphi \left( z_1 \over z_0 \right)\;.
\label{recursion}
 \end{eqnarray}
 As we have hypothesized that (\ref{main-claimbis}) is valid for $p_0 \leq P$,
we can replace each term of (\ref{recursion}) by the corresponding
expressions (\ref{rhs}), leading to
 \begin{eqnarray}
\mbox{l.h.s. of (\ref{main-claimbis})}
  &=& \sum_{m_0=0}^\infty
\sum_{m_1=0}^\infty
{c_{m_0+P} \over
  c_{m_0} c_{m_1}} \  {z_1^{m_1} \over z_0^{m_0}} \times \nonumber\\ 
 &&\hskip15pt \left[  (1 - q^{p_1}) \delta_{m_0+P, m_1+p_1-1}
     +  z_1 \ q^{p_1} \ \delta_{m_0+P, m_1+p_1} \right]
\nonumber\\
 &=& \sum_{m_0=0}^\infty 
\sum_{m_1=0}^\infty
{c_{m_0+P+1} \over
  c_{m_0} c_{m_1}} \  {z_1^{m_1} \over z_0^{m_0}}  \ 
 \delta_{m_0+P+1, m_1+p_1-1}\;,
 \end{eqnarray}
 which is identical for $p_0=P+1$ to (\ref{rhs}).  This completes the
derivation of (\ref{main-claimbis}). 

\subsection{Analytic continuation of (\ref{main-claimbis})\label{AC}}

 Recall that (\ref{main-claimbis}) has been established for $|z_0| > |z_1|$,
with integration contour a circle $|z|=R$ with $|z_1|<R<|z_0|$.  However, the
left hand side of this equation is an analytic function defined for all
complex $z_0$ and $z_1$ except at $z_0=0$ and at the poles of
$\varphi(z_1/z_0)$.  In this section we make an analytic continuation
of the right hand side to obtain an integral representation of the left side
which is valid for any $z_0$, $z_1$ for which the left side is defined. 
In the final representation we will again
integrate over a contour $|z|=R$, but now
$R$ will be allowed to take any value for which the contour does not pass
through singularities of the integrand, that is, for which
 \begin{equation}
 \label{goodR}
 R>0,\qquad R\ne q^m|z_1|,\ m\ge0, \qquad R\ne q^{-m}|z_0|,\ m\ge0.
 \end{equation}

The expression on the right hand side of (\ref{main-claimbis}) must be
modified when, as one varies $z_0$, $z_1$ and $R$, the singularities of the
integrand---that is, the poles of $\varphi(q^nz/z_0)$ and
$\varphi(z_1/z)$---cross the integration contour.  When this happens,
however, the residue theorem tells us how (\ref{main-claimbis}) is to be
modified: one simply includes the residue of the pole on the right hand side
of the equation, adding or subtracting it according to whether the pole
crosses the contour from the inside to the outside or vice versa.  Using the
fact that the residue of $\varphi(z)$ at $z=q^{-m}$, $m=0,1,\ldots$, is
 \begin{equation}
   {(-1)^{m+1}q^{m(m-1)/2}\over c_m}\varphi(q),
 \end{equation}
 one then finds the following extension of (\ref{main-claimbis}): 

Suppose that $R$ satisfies (\ref{goodR}).  If $R<|z_1|$ then define $k_1$ by
\begin{equation}
 q^{k_1+1} |z_1| < R < q^{k_1} |z_1|\;,
\label{conditionk1}
\end{equation}
and if $R>|z_0|$ define $k_0$ by 
\begin{equation}
      {|z_0|\over q^{k_0}} < R < {|z_0| \over q^{k_0 +1}}\;.
\label{conditionx}
\end{equation}
 Then 
 \begin{eqnarray}
 {\langle z_0|X_0X_1|z_1\rangle\over\langle z_0|z_1\rangle}
 \varphi \left( z_1 \over z_0 \right)&&\nonumber\\
 &&\hskip-100pt=\;  
   \sum_{n=0}^\infty  {q^n \over c_n}
  \Biggl\{{1\over\varphi(q)}
   \oint_{|z|=R} {dz \over 2 \pi i z}
 {\langle z_0|X_0|q^nz\rangle\over\langle z_0|q^nz\rangle}
 \varphi \left( q^n z \over z_0 \right)
 {\langle z|X_1|z_1\rangle\over\langle z|z_1\rangle}
  \varphi \left( z_1 \over z \right) 
 \nonumber\\
 &&\hskip-60pt+\;\Theta(|z_1|-R)
  \sum_{m=0}^{k_1} {(-1)^mq^{m(m+1)/2}\over c_m}
  \varphi\left(q^{n+m}z_1\over z_0\right)\times
   \nonumber\\
   && 
 {\langle z_0|X_0|q^{n+m}z_1\rangle\over\langle z_0|q^{n+m}z_1\rangle}\,
 {\langle q^mz_1|X_1|z_1\rangle\over\langle q^mz_1|z_1\rangle}
 \nonumber\\
&&\hskip-60pt+\;\Theta(R-q^{-n}|z_0|)
   \sum_{m=0}^{k_0-n} {(-1)^mq^{m(m+1)/2}\over c_m}
  \varphi\left(q^{n+m}z_1\over z_0\right)\times
    \nonumber\\
    && 
 {\langle z_0|X_0|q^{-m}z_0\rangle\over\langle z_0|q^{-m}z_0\rangle}\,
 {\langle q^{-(n+m)}z_0|X_1|z_1\rangle\over\langle q^{-(n+m)}z_0|z_1\rangle}
 \label{main-claim4}
 \end{eqnarray}

\subsection{Asymptotics of the additivity formula\label{asymptotics}}

We now ask what happens to the representation (\ref{main-claim4}) when the
system size $N$ becomes very large.  Throughout this section we take
$z_0$ and $z_1$ to be positive real numbers  and $X_0$ and $X_1$ to be
polynomials, with positive coefficients, in the operators $D$ and $E$.  We
will use throughout this section three properties of the product
 \begin{equation}
 \label{product}
\Pi_n(z)\equiv    {\langle z_0|X_0|q^nz\rangle\over\langle z_0|q^nz\rangle}\;
    {\langle z|X_1|z_1\rangle\over\langle z|z_1\rangle}
 \end{equation}
 occurring in the integrands of the representation (\ref{main-claim4}), which
follow from the discussion of Section~\ref{prelims}: $\Pi_n(z)$ is a
polynomial in $q^nz$ and $1/z$, with positive coefficients; for $z$ on the
positive real axis, $\Pi_n(z)$ is a convex function of $z$; for fixed
positive $z$, $\Pi_n(z)$ decreases as $n$ increases.  We will assume that,
for $z$ on the positive axis, $\Pi_n(z)$ grows exponentially in $N$. 

Now $\Pi_0(z)$ has a unique minimum at some value $z_{\rm min}$ on the
positive real axis; we will use the representation (\ref{main-claim4}) with
the choice $R=z_{\rm min}$.  Using first the fact that the maximum of the
magnitude $|\Pi_n(z)|$ on the contour $|z|=z_{\rm min}$ occurs on the real
axis, and then the monotonicity of $\Pi_n(z)$ in $n$ for $z$ positive, we see
that each integral occurring in (\ref{main-claim4}) satisfies
 \begin{eqnarray}
   \label{saddle}
   \oint_{|z|=z_{\rm min}} {dz \over 2 \pi i z}
    {\langle z_0|X_0|q^nz\rangle\over\langle z_0|q^nz\rangle}
      \varphi \left( q^nz \over z_0 \right)
 {\langle z|X_1|z_1\rangle\over\langle z|z_1\rangle}
     \varphi \left( z_1 \over z \right) 
   \hskip-150pt\nonumber&&\\
  &\lesssim&  
    {\langle z_0|X_0|q^nz_{\rm min}\rangle
           \over\langle z_0|q^nz_{\rm min}\rangle}
    {\langle z_{\rm min}|X_1|z_1\rangle
             \over\langle z_{\rm min}|z_1\rangle}\;,\nonumber\\
  &\le&  
    {\langle z_0|X_0|z_{\rm min}\rangle
           \over\langle z_0|z_{\rm min}\rangle}
    {\langle z_{\rm min}|X_1|z_1\rangle
             \over\langle z_{\rm min}|z_1\rangle}\;,
 \end{eqnarray}
 where the first inequality holds up to factors that do not grow exponentially
with $N$.  On the other hand, when $n=0$ the point $z_{\rm min}$ will be a
saddle point for $\Pi_0(z)$ lying on the contour $|z|=z_{\rm min}$, and
equality (again up to factors not growing exponentially with $N$) will hold
in (\ref{saddle}). 

The bound (\ref{saddle}), and an argument similar to that above for the terms
in (\ref{main-claim4}) arising from the poles, show that the $n=0$ terms
there always dominate those with $n>0$, so that we may neglect the latter. 
But for $n=0$, the fact that $\Pi_0(z)$ is convex, with a minimum at $z_{\rm
min}$, implies that the $m=0$ terms dominate each sum over $m$.  Thus
(\ref{main-claim4}) becomes
 \begin{eqnarray}
 {\langle z_0|X_0X_1|z_1\rangle\over\langle z_0|z_1\rangle}
  &\simeq& \max\Biggl\{ 
    {\langle z_0|X_0|z_{\rm min}\rangle
           \over\langle z_0|z_{\rm min}\rangle}
    {\langle z_{\rm min}|X_1|z_1\rangle
             \over\langle z_{\rm min}|z_1\rangle}\;,
    \nonumber\\
   &&\;\Theta(z_1-z_{\rm min})
 {\langle z_0|X_0|z_1\rangle\over\langle z_0|z_1\rangle}\,
 {\langle z_1|X_1|z_1\rangle\over\langle z_1|z_1\rangle}\;,
   \nonumber\\
   &&\;\Theta(z_{\rm min}-z_0)
 {\langle z_0|X_0|z_0\rangle\over\langle z_0|z_0\rangle}\,
 {\langle z_0|X_1|z_1\rangle\over\langle z_0|z_1\rangle}
  \Biggl\}\;.
 \label{intermediate}
 \end{eqnarray}
 Now the discussion can be completed by considering successively all the
possible relative positions of $z_0$, $z_1$ and $z_{\rm min}$. 

 Suppose first that $z_1<z_0$.  Then  (\ref{intermediate}) gives 
 \begin{eqnarray}
 {\langle z_0|X_0X_1|z_1\rangle\over\langle z_0|z_1\rangle}
    \simeq\left\{\begin{array}{ll}
 \displaystyle  {\langle z_0|X_0|z_{\rm min}\rangle
           \over\langle z_0|z_{\rm min}\rangle}
    {\langle z_{\rm min}|X_1|z_1\rangle
             \over\langle z_{\rm min}|z_1\rangle}, &
               \mbox{if $z_1<z_{\rm min}<z_0$},\\ \noalign{\vskip3pt}
 \displaystyle {\langle z_0|X_0|z_1\rangle\over\langle z_0|z_1\rangle}\,
 {\langle z_1|X_1|z_1\rangle\over\langle z_1|z_1\rangle}, &
               \mbox{if $z_{\rm min}<z_1<z_0$},\\ \noalign{\vskip3pt}
 \displaystyle {\langle z_0|X_0|z_0\rangle\over\langle z_0|z_0\rangle}\,
 {\langle z_0|X_1|z_1\rangle\over\langle z_0|z_1\rangle}, &
               \mbox{if $z_1<z_0<z_{\rm min}$},
   \end{array}\right.
 \label{Wfan1}
 \end{eqnarray}
 and one sees that each case in (\ref{Wfan1}) reduces to
 \begin{equation}
 {\langle z_0|X_0X_1|z_1\rangle\over\langle z_0|z_1\rangle}
    \simeq\min_{z_1\le z\le z_0}
 {\langle z_0|X_0|z\rangle\over\langle z_0|z\rangle}\,
 {\langle z|X_1|z_1\rangle\over\langle z|z_1\rangle}.
 \label{Wfan}
 \end{equation}
 On the other hand, if $z_0<z_1$,  then  (\ref{intermediate}) gives 
 \begin{eqnarray}
 {\langle z_0|X_0X_1|z_1\rangle\over\langle z_0|z_1\rangle}
    \simeq\left\{\begin{array}{ll}
     \max\limits_{z=z_0,z_1}
 \displaystyle  {\langle z_0|X_0|z\rangle
           \over\langle z_0|z\rangle}
    {\langle z|X_1|z_1\rangle
             \over\langle z|z_1\rangle}, &
               \mbox{if $z_0<z<z_1$},\\ \noalign{\vskip3pt}
 \displaystyle {\langle z_0|X_0|z_1\rangle\over\langle z_0|z_1\rangle}\,
 {\langle z_1|X_1|z_1\rangle\over\langle z_1|z_1\rangle}, &
               \mbox{if $z_{\rm min}<z_0<z_1$},\\ \noalign{\vskip3pt}
 \displaystyle {\langle z_0|X_0|z_0\rangle\over\langle z_0|z_0\rangle}\,
 {\langle z_0|X_1|z_1\rangle\over\langle z_0|z_1\rangle}, &
               \mbox{if $z_0<z_1<z_{\rm min}$},
   \end{array}\right.
 \label{Wshock1}
 \end{eqnarray}
 and each case in (\ref{Wshock1}) reduces to
 \begin{equation}
 {\langle z_0|X_0X_1|z_1\rangle\over\langle z_0|z_1\rangle}
    \simeq\max_{z=z_0,z_1}
   {\langle z_0|X_0|z\rangle \over\langle z_0|z\rangle}
   {\langle z|X_1|z_1\rangle \over\langle z|z_1\rangle}.
 \label{Wshock}
 \end{equation}

\subsection{Derivation of  (\ref{additivity-asym1}) and 
   (\ref{additivity-asym2})\label{finalstep}}

 We finally want to obtain the fundamental additivity relations for
${\cal H}$, (\ref{additivity-asym1}) and (\ref{additivity-asym2}), from
(\ref{Wfan}) and (\ref{Wshock}).  The first step is to relate the densities
$\rho_a,\rho_b$ to the parameters $z_0,z_1$.  Using (\ref{reservoir-asym}),
(\ref{vdef}) and (\ref{wdef}), one can
easily establish that if
 \begin{equation}
 \label{rhoz}
 \rho_a = {z_0 \over 1 + z_0} \;, \qquad \rho_b = {z_1 \over 1 + z_1}\;,
 \end{equation}
 then 
 \begin{equation}
  \label{WW}
 {\langle z_0|X_0X_1|z_1\rangle\over\langle z_0|z_1\rangle}
 = {\langle W | X_0X_1 | V \rangle \over\langle W | V \rangle }\;,
 \end{equation}
 where $\langle W|$ and $|V\rangle$ are defined by 
(\ref{alg2}) and (\ref{alg3}).

 Now let us consider a given profile $\rho(x)$ defined for $a < x < b$, and
for fixed $c$ with $a<c<b$ denote by $X_0$ the sum over all the products of
$D$'s and $E$'s consistent with the left part of this profile over the first
$L(c-a) $ sites, and by $X_1$ the same quantity for the right part of the
profile over the last $L(b-c)$ sites.  We define ${\cal H}$ by
 \begin{equation}
 \label{XHconnection}
  (1-q)^{L(b-a)}
 {\langle z_0|X_0X_1|z_1\rangle\over\langle z_0|z_1\rangle}
   \sim \exp \left[ - L{\cal H}_{[a,b]} (\{\rho(x)\};\rho_a,\rho_b)\right]\;.
 \end{equation}
 Then we obtain immediately (\ref{additivity-asym1}) and
(\ref{additivity-asym2}) from
(\ref{Wfan}) and (\ref{Wshock}).  Moreover from
(\ref{calFdef}), (\ref{matrix}), (\ref{WW}), and (\ref{XHconnection}) 
we see that the constant
$K(\rho_a,\rho_b)$ which appears in (\ref{Hdef-asym}) is given by
\begin{equation}
  \label{Kexpression}
  (1-q)^L(b-a) \langle W | (D+E)^L(b-a) | V \rangle 
   \sim e^{- L(b-a) K(\rho_a,\rho_b)}\;.
 \end{equation}
  Writing (\ref{Wfan}) and (\ref{Wshock}) 
for  $X_0= (D+E)^{L(c-a)}$
and $X_1= (D+E)^{L(b-c})$, we see that $K(\rho_a,\rho_b)$ should satisfy
 \begin{equation}
 (b-a) K(\rho_a,\rho_b) = 
 \sup_{\rho_b \leq \rho_c \leq \rho_a} \left[ (c-a) K(\rho_a,\rho_c)
    + (b-c) K(\rho_c,\rho_b) \right]\;,
 \label{K1}
 \end{equation}
 if $\rho_a > \rho_b$, and
 \begin{equation}
  (b-a) K(\rho_a,\rho_b) = 
  \min_{\mbox{$\rho_c=\rho_a$ or $\rho_b$}}  
   \left[ (c-a) K(\rho_a,\rho_c) + (b-c) K(\rho_c,\rho_b) \right]\;,
 \label{K2}
 \end{equation}
 if $\rho_a < \rho_b$.  Now when $\rho_a=\rho_b$ the matrices $D$ and $E$
commute and may be realized as the scalars $D=(1-q)^{-1}(1+z_0)$, 
$E=(1-q)^{-1}(1+z_0^{-1})$, so that 
from (\ref{Kexpression}) and (\ref{rhoz}) we have 
 \begin{equation}
  K(\rho_a,\rho_a)= - \log \left( z_0 +2 + {1 \over z_0} \right) 
   = \log [\rho_a (1- \rho_a) ]\;.
 \label{K3}
 \end{equation}
 From (\ref{K1}) and (\ref{K3}) one finds, by repeated subdivision of the
interval $[a,b]$, that if $\rho_a>\rho_b$ then
 \begin{equation}
 (b-a) K(\rho_a,\rho_b) =  \sup_f\int_a^b dx\,\log[f(x)(1-f(x))]\;,
 \label{K1a}
 \end{equation}
 where the supremum is over nonincreasing functions $f(x)$ with
$f(a)=\rho_a$ and $f(b)=\rho_b$, and  from (\ref{K1a}) one obtains
(\ref{Kdef1}).  Similarly, if $\rho_a<\rho_b$ one obtains from 
(\ref{K1}) and (\ref{K3}) that 
 \begin{equation}
 (b-a) K(\rho_a,\rho_b) =  \inf_{a\le y\le b}
 \left\{\int_a^y dx\,\log[\rho_a(1-\rho_a)]
    +\int_y^b dx\,\log[\rho_b(1-\rho_b)]\right\}\;,
 \label{K2a}
 \end{equation}
 and (\ref{Kdef2}) follows.

\section{Large deviations versus typical fluctuations\label{LD_Fluctuations}}

Expressions (\ref{result}) and (\ref{result3}) enable us to calculate the
large deviation function for an arbitrary density profile $\rho(x)$.  As we
have already noted and will show in this section, the large deviation
functional, which describes macroscopic (order $N$) deviations from the
typical profile $\bar\rho$, is in region $C$, where
 \begin{equation}
\rho_a > {1 \over 2} > \rho_b \;,
 \end{equation}
   not simply related to the fluctuations
(order $\sqrt N$) around $\bar\rho$.  We demonstrate this by computing
explicitly, for $q=0$, the probability of seeing a given global density $r$
in a window $c<x<d$ of our system, with no other constraint in the system,
both for fixed $r$ with $r\ne\bar\rho$ and for $r-\bar\rho$ of order $1/\sqrt
N$. 

 We divide our system of $N=L(b-a)$ sites into three boxes, of $N_1=L(c-a)$
sites, $N_2= L(d-c) $ sites and $N_3=N (b-d)$ sites, corresponding to
macroscopic intervals $[a,c]$, $[c,d]$, and $[d,b]$, and compute the
probability that the density is $r$ in the middle box, i.e., the probability
$P(M_2)$  that the total
number $M_2$ of particles in the middle box is
 \begin{equation}
  M_2 = L(d-c) r,
 \label{density}
 \end{equation}
 with no constraint imposed in the two other boxes. 

\subsection{Large deviation\label{Discontinuous_LD}}

Corresponding to the above constraint there will be an optimal profile
$\rho(x)$ for which, with
 $\bar{\cal F}(r)\equiv{\cal F}_{[a,b]}(\rho(x);\rho_a,\rho_b)$, one has 
 \begin{equation}
   P(M_2)\sim\exp[-L\bar{\cal F}].\label{PM2}
 \end{equation}
 Since $\rho_a> \rho_b$, one can use the additivity formula
(\ref{additivity-asym1}) 
 \begin{eqnarray}
  {\cal H}_{[a,b]}(\rho(x);\rho_a,\rho_b) 
 = \sup_{\rho_b \leq \rho_d \leq \rho_c \leq \rho_a}
  \left\{{\cal H}_{[a,c]}(\rho(x);\rho_a,\rho_c)\right. && \nonumber\\
  &&\hskip-180pt  \label{add1}
   \left.{} + {\cal H}_{[c,d]}(\rho(x);\rho_c,\rho_d) 
+ \{{\cal H}_{[d,b]}(\rho(x);\rho_d,\rho_b)  \right\}.
  \end{eqnarray}
 As there is no constraint on the profile in the two side boxes, one has
 ${\cal F}_{[a,c]}(\rho(x);\rho_a,\rho_c)
 ={\cal F}_{[d,b]}(\rho(x);\rho_d,\rho_b)=0$, so that from (\ref{Hdef-asym}),
 \begin{eqnarray}
  {\cal H}_{[a,c]}(\rho(x);\rho_a,\rho_c)&=& (c-a) K(\rho_a,\rho_c),
    \label{Hleft}\\
  {\cal H}_{[b,d]}(\rho(x);\rho_d,\rho_b)&=& (b-d) K(\rho_d,\rho_b).
  \label{Hright}
 \end{eqnarray}
   Moreover, since $\rho_c \geq \rho_d$ we know from Sections~\ref{flat} and
\ref{number} that within the central box $[c,d]$ the optimal
profile $\rho(x)$, which corresponds to a fixed number of particles there,
is flat.  From  (\ref{Hdef-asym}) and (\ref{flat1}) we then have 
 \begin{equation}
 {\cal H}_{[c,d]}(\rho(x);\rho_c,\rho_d)
 = \left\{
    \begin{array}{ll}
      (c-d)[r\log(r(1-\rho_c)+(1-r\log((1-r)\rho_c)],\hskip-20pt \\
   \noalign{\vskip3pt}
      (c-d)[2r\log r+2(1-r\log(1-r)],\\
   \noalign{\vskip3pt}
      (c-d)[r\log(r(1-\rho_d)+(1-r\log((1-r)\rho_d)],\hskip-20pt\\
    \end{array}
\right.
 \label{flat1bis}
 \end{equation}
 when $r<1-\rho_c$, $1-\rho_c<r<1-\rho_d$, and $1-\rho_d<r$, respectively. 
 Now from (\ref{add1}) one must choose $\rho_c$ and $\rho_d$ to maximize the
sum of (\ref{Hleft}), (\ref{Hright}), and (\ref{flat1bis}).  The result
depends on the sign of $r-1/2$.

 \smallskip\noindent
 {\bf Case 1: $r<1/2$.} In this case the optimizing values of $\rho_c$ and
$\rho_d$ are
 \begin{equation}
  \label{rhoc}
   \rho_c = \min \left\{ { (d-a) - r (d-c) \over d+c - 2 a} , \rho_a\right\}
  \;,   \qquad  \rho_d = {1\over 2}.
\end{equation}
 The corresponding LDF is 
 \begin{eqnarray}
   {\bar{\cal F}(r)} 
   &=&  (d-c) [r\log(4r(1 - \rho_c))  
  + (1-r) \log (4(1-r)\rho_c) ] \nonumber\\
    && \hskip30pt  {} + (c-a)\log(4\rho_c(1-\rho_c)) ,
    \label{F1}
 \end{eqnarray}
 which for $r$ close to $1/2$ becomes
 \begin{equation}
 \label{F1bis}
  {\bar{\cal F}(r)}
  = {4 (d-c) (d-a) \over d+c - 2 a}\left( r - {1 \over 2} \right)^2 
   + O  \left( r - {1 \over 2} \right)^3\;.
\end{equation}
 We also find from Section~\ref{MLP} that the optimal profile $\rho(x)$
satisfies $\rho(x)= \rho_c$ for $a < x < c$ and
$\rho(x)= 1/2$ for $d < x< b$. 

 \smallskip\noindent
 {\bf Case 2: $r>1/2$.} In this case, a similar calculation leads to
\begin{equation}
  \label{rhod}
   \rho_c = {1 \over 2} \;,  \qquad  
    \rho_d = \max \left\{ { (b-c) - r (d-c) \over 2b-d -c} , \rho_b\right\}  
\end{equation}
and 
 \begin{eqnarray}
  {\bar{\cal F}(r)} &=&   
     (d-c) [r\log(4r(1 - \rho_d))  + (1-r)\log(4(1-r) \rho_d)] \nonumber\\
  &&\hskip30pt {} + (b-d) \log( 4 \rho_d (1-\rho_d)); \label{F2} 
 \end{eqnarray}
  and for $r$ close to $1/2$, (\ref{F2}) gives
 \begin{equation}
 \label{F2bis}
  {\bar{\cal F}} 
  =  {4 (d-c) (b-c)\over 2 b - c - d}\left( r - {1 \over 2}\right)^2 
   + O  \left( r - {1 \over 2} \right)^3.
\end{equation}

 We see from (\ref{F1bis}) and (\ref{F2bis}) that in general the limiting
value of the second derivative of $\bar{\cal F}(r)$ at $r=1/2$ depends on
the sign of $r-1/2$, so that ${\bar{\cal F}(r)}$ is in general nonanalytic
at $r=1/2$. Note, however, that when the overall density for the system is
specified, i.e., when $c=a$ and $d=b$, $\bar{\cal F}(r)$ is analytic.

\subsection{Fluctuations\label{fluct_vs_LD}}

Formulae (\ref{F1}) and (\ref{F2}), with (\ref{rhoc}) and (\ref{rhod}),
give us the leading behavior of $P(M_2)$
for large deviations, i.e. for $r-1/2$ of order $1$.
Let us now discuss the small fluctuations, i.e., 
the regime in which  $r-1/2 = O(L^{-1/2})$, so that $L{\cal F}$ is of order
one.  As (\ref{F1bis}) and (\ref{F2bis}) 
do not coincide, one expects that typical 
fluctuations around the optimal profile
 $\bar\rho(x)=1/2$ will be anomalous, i.e, non-Gaussian.
Let us define the fluctuation $\mu$  in the number $M_2$ of particles
in the  central box by
\begin{equation}
\label{mudef}
M_2 - {(d-c)L \over 2} =  \mu \sqrt{L} .
\end{equation}
 In the next subsection we will derive, for $q=0$, the following probability
density $p(\mu)= P(M_2) \sqrt{L}$ for the random variable $\mu$:
 \begin{eqnarray}
  \label{pofmu}
 p(\mu) 
   &=&  {8 \ (b-a)^{3/2}\over[\pi (c-a)(b-d) ]^{3/2} \ (d-c) }
  \times \\ 
  && \hskip-20pt\int_0^\infty dx \int_0^\infty dy \ x y\,
   \exp \left[ - {x^2 \over c-a} - {y^2 \over b - d}\right] \times
   \nonumber \\
 && \hskip-20pt
  \left\{ \exp \left[ - 2 \ { \mu^2 + (\mu + x - y)^2 \over d-c} \right] 
 - \exp \left[ - 2 \ { (\mu+x)^2 + (\mu - y)^2 \over d-c} \right] \right\}.
 \nonumber
 \end{eqnarray}
 We see that indeed $\mu$ has a non-Gaussian distribution.
Note, however, that the $\sqrt L$ scaling in
(\ref{mudef}) is that appropriate for ``normal'' fluctuations. 

 For large positive $\mu$, (\ref{pofmu}) becomes
 \begin{eqnarray}
   p(\mu)&=& { 4 \over \mu \pi}
   {(b-a)^{3/2}(b-c) ( b-d) 
     \over (d-c)^{1/2} (2 b - c -d)^{3/2} (c-a)^{3/2} }\times
   \nonumber \\
 &&\hskip80pt \exp \left[ - {4 (b-c) \over (d-c) (2b - d - c) } \mu^2 \right],
  \label{mularge1}
 \end{eqnarray}
 and for large negative $\mu$, 
 \begin{eqnarray}
   p(\mu)&=& {  4 \over |\mu| \pi}
  {(b-a)^{3/2}(d-a) ( c-a) 
      \over (d-c)^{1/2} (d+c - 2 a)^{3/2} (b-d)^{3/2} }\times
  \nonumber\\
  &&\hskip80pt \exp\left[-{4 (d-a) \over (d-c) ( d + c - 2 a) } \mu^2 \right].
   \label{mularge2}
 \end{eqnarray}
 We see that the large $\mu$ asymptotics (\ref{mularge1}) and
(\ref{mularge2}) match with (\ref{F1bis}) and (\ref{F2bis}) when
$|\mu|\gg1$ and $|r-1/2|\ll1$, with $\mu= (r-1/2) \sqrt{L}$, i.e., that
(\ref{pofmu}) interpolates between the large deviation regions $r>1/2$ and
$ r < 1/2$. A similar relationship has been found between the distributions
of large deviations and of typical fluctuations of the current in the ASEP
on a ring \cite{Varadhan}.

Also, from (\ref{pofmu}) one can compute 
the average $\langle \mu \rangle$ of $\mu$:
 \begin{equation}
   \langle \mu \rangle 
    = \int p(\mu) \mu d \mu 
    = { \sqrt{(d-a)(b-d)} - \sqrt{(c-a)(b-c)} \over \sqrt{\pi}\sqrt{b-a}}\;.
 \label{muaverage}
 \end{equation}
 For a system of $N=L(b-a)$ sites, one sees from (\ref{tauiasymp})
that
\begin{equation}
 \sum_{i=L(c-a)}^{L(d-a)} \left(\langle \tau_i \rangle - {1 \over 2}\right)
   = {\sqrt{L} \over 2 \sqrt{\pi}}
  \int_c^d {b+a - 2 x \over \sqrt{(b-a)(b-x)(x-a)}} dx,
 \end{equation}
in agreement with (\ref{muaverage}), since
 $\langle M_2 \rangle - L (d-c)/2 = \langle \mu \rangle \sqrt{L} $. 

\subsection{Derivation of (\ref{pofmu})\label{nongaussian}}

One way to derive (\ref{pofmu}) is to use the following explicit
 representation \cite{DEHP} of (\ref{alg1}), valid for $q=0$:
\begin{eqnarray}
\label{Drep}
  D &=& \sum_{n=1}^\infty | n \rangle \langle n | + |n \rangle \langle n+1|,
   \\
  \label{Erep}
  E &=& \sum_{n=1}^\infty | n \rangle \langle n | + |n+1 \rangle \langle n|,
\end{eqnarray}
 where the vectors $|1 \rangle, |2 \rangle, ...|n \rangle ...$ form an
orthonormal basis of an infinite dimensional space (with $\langle n | m
\rangle = \delta_{n,m}$).  Within this basis, the vectors $|V \rangle$ and
$\langle W|$ are given (see (\ref{alg2}), (\ref{alg3}), and
(\ref{reservoir-asym})) by
\begin{eqnarray}
\label{Vrep}
  |V \rangle 
  = \sum_{n=1}^\infty  \left( 1 - \beta \over \beta \right)^n | n \rangle 
  = \sum_{n=1}^\infty  \left( \rho_b \over 1 - \rho_b \right)^n | n \rangle,
  \\
 \label{Wrep}
 \langle W | 
  = \sum_{n=1}^\infty  \left( 1 - \alpha \over \alpha \right)^n \langle  n  |
  = \sum_{n=1}^\infty  \left( 1- \rho_a \over  \rho_a \right)^n  \langle n |,
\end{eqnarray}
 and one can show (for example by recursion) that
 \begin{equation}
  \langle p | (D+E)^N | p' \rangle = { (2N)! \over (N+p-p')! \  (N+p'-p)! }
- { (2N)! \over (N+p+p')!  \ (N-p'-p)! }
\label{DplusE}
 \end{equation}
and that \cite{Mal}
\begin{eqnarray}
  \langle p| X_{N,M} | p' \rangle =
  { (N!)^2 \over (M)! \  (N-M)! \  (M+p-p')! \  (N-M-p+p')!}
\nonumber \\
- { (N!)^2 \over (M+p)! \  (N-M-p)! \  (M-p')! \   (N-M+p')!}
\label{XNM}
 \end{eqnarray}
 where $X_{N,M}$ is the sum over all the configurations of $N$ sites with $M$
occupied particles.  The probability $P(M_2)$ that the number of particles is
$M_2$ in the central box is given by
\begin{equation}
  P(M_2) = {\langle W | (D+E)^{N_1} \  X_{N_2,M_2} \  (D+E)^{N_3} | V \rangle
 \over \langle W | (D+E)^{N_1+N_2+N_3} | V \rangle }\;.
 \label{PM2bis}
 \end{equation}

Let us first analyze the denominator of (\ref{PM2bis}):
 \begin{equation}
 \langle W | (D+E)^{N} | V \rangle  = \sum_{p_1=1}^\infty
  \sum_{p_2=1}^\infty 
 \left( 1 - \rho_a \over \rho_a \right)^{p_1}
 \left( \rho_b \over 1 - \rho_b \right)^{p_2} 
\langle p_1 | (D+E)^{N} | p_2 \rangle .
 \end{equation}
 For large $N$, this sum is dominated by $p_1$ and $p_2$ of order 1,
so one can use an approximation of (\ref{DplusE}) 
valid for $p$ and $p'$ of order  $\sqrt{N}$ or less,
 \begin{equation}
\langle p | (D+E)^N | p' \rangle \simeq { 4^N \over \sqrt{\pi N} }
\left[ \exp \left(-(p-p')^2 \over N \right)
- \exp \left(-(p+p')^2 \over N \right) \right],
\label{DplusEbis}
\end{equation}
 and one gets for large $N$,
 \begin{equation}
 \langle W | (D+E)^{N} | V \rangle  \simeq  {4^{N+1} \over \sqrt{\pi} N^{3/2}}
\ {(1- \rho_a) \rho_a \over (2 \rho_a -1)^2 }
\ {(1- \rho_b) \rho_b \over (2 \rho_b -1)^2 } .
\label{denom-asympt}
 \end{equation}
One can write the numerator of (\ref{PM2bis}) as
 \begin{eqnarray}
 \langle W | (D+E)^{N_1} \  X_{N_2,M_2} \  (D+E)^{N_3} | V \rangle
= \sum_{p_1} \sum_{p_2} \sum_{p_3} \sum_{p_4}
 \langle W | p_1 \rangle
\nonumber \\ 
 \langle p_1| (D+E)^{N_1} |  p_2 \rangle 
   \langle p_2| X_{N_2,M_2}|  p_3 \rangle 
   \langle p_3 | (D+E)^{N_3}| p_4 \rangle \langle p_4 | V \rangle.
\label{numerator}
\end{eqnarray}
 When $N_1,N_2,N_3$ are large and of order $L$ and when the difference $M_2 -
N_2/2$ is of order $\sqrt{L}$ as in (\ref{mudef}), these sums are dominated
by $p_1$ and $p_4$ of order $1$ and $p_2$ and $p_3$ of order $\sqrt{L}$.  If
one writes
 \begin{equation}
  p_2 =  x \sqrt{L},  \qquad  p_3 = y \sqrt{L},
 \end{equation}
 one gets that
 \begin{eqnarray}
 \langle p_2| X_{N_2,M_2}|  p_3 \rangle
  &\simeq& 4^{N_2} {2 \over \pi  N_2} 
  \left[ \exp \left( -{2 L \over N_2} (\mu^2 + (\mu+x-y)^2)\right)\right.
   \nonumber\\
 &&\hskip20pt {} - 
  \left.\exp \left( -{2 L \over N_2} ((\mu-y)^2 + (\mu+x)^2) \right) \right],
 \end{eqnarray}
 and the numerator becomes, after summing over $p_1$ and $p_4$ and
replacing the sums over $p_2$ and $p_3$ by integrals
 \begin{eqnarray}
    \langle W | (D+E)^{N_1} \  X_{N_2,M_2} \  (D+E)^{N_3} | V \rangle
  &=&\label{numeratorbis}\\
 &&\hskip-181pt 
  {4^{N_1+N_2+N_3+2} \times 2 \over \pi^2 L^2 [(c-a)(b-d)]^{3/2} (d-c)} 
   \times
\nonumber \\ 
  &&\hskip-171pt   {(1- \rho_a) \rho_a \over (2 \rho_a -1)^2 }
\ {(1- \rho_b) \rho_b \over (2 \rho_b -1)^2 } 
\
 \int_0^\infty dx
\int_0^\infty dy \  x  \ y \ \exp \left[  - {x^2 \over c-a} - {y^2 \over b - d}
\right]  \times
\nonumber \\
  &&\hskip-171pt  \left\{  \exp \left[   - 2 \ { \mu^2 + (\mu + x - y)^2 \over d-c} \right]
- \exp \left[   - 2 \ { (\mu+x)^2  + (\mu - y)^2 \over d-c} \right] \right\}
 \nonumber
\end{eqnarray}
 which reduces to (\ref{pofmu}) after dividing by  (\ref{denom-asympt}).

{\bf Remark}: One can also recover from (\ref{PM2bis}) the expressions
(\ref{F1}) and (\ref{F2}) by allowing deviations in $M_2$ of order $L$.

\section{Conclusion\label{conclusion}}

The main results of the present work are the exact expressions (\ref{result})
and (\ref{result3}) for the LDF ${\cal F}(\{\rho\})$ for the SNS of the open
ASEP in one dimension and the simple additivity formulae
(\ref{additivity-asym1}) and (\ref{additivity-asym2}) that they satisfy. 

  This ${\cal F}$, like the one we found for the symmetric case in
\cite{DLS,DLS2}, is a non-local functional of the density profile
$\{\rho(x)\}$.  We expect non-locality to be a general feature of such
functionals for non-equilibrium systems. 

Our expressions of the  LDF, which take different forms in the fan region
$\rho_a> \rho_b$ (where the reservoirs and the bulk asymmetry cooperate)
and in the shock region $\rho_a< \rho_b$ (where they act in opposite directions), reflect several qualitative differences:

In the fan region, $\rho_a > \rho_b$, the probability of macroscopic
deviations from the typical density profile is reduced compared with that in
an equilibrium system with the same typical profile (see (\ref{suppress}));
this was also true in the symmetric case \cite{DLS}.  Another surprising
feature of the fan region, at least in the maximal current phase C, is that
the fluctuations of the density profile cannot be calculated from the LDF,
and that these fluctuations are in general not Gaussian (see section 6).  We
have no heuristic explanation of this behavior. 

 In the shock region of the phase diagram, $\rho_a < \rho_b$, ${\cal F}$ is
not convex in $\rho$ (see section 3.3).  Moreover, in this region the
probability of macroscopic deviations from typical behavior is increased
rather than reduced; see (\ref{enhance}).  This enhancement of deviations
appears similar to known behavior \cite{Exp} of fluctuations (the behavior of
macroscopic deviations is not known) in a slab of fluid in contact at the top
with a heat reservoir at temperature $T_a$, and at the bottom with another
reservoir at temperature $T_b$: the Rayleigh-B{\'e}nard system \cite{CH}.  In
this system the force of gravity causes the SNS to undergo dynamic phase
transitions when $(T_b - T_a)$ is ``sufficiently'' large, corresponding to
different spatial patterns of heat and mass flow.  This transition is
preceded, as $T_b-T_a$ is increased, by enhanced fluctuations even for very
small differences between the two temperatures (for which the system is
stable) as long as $T_b>T_a$. 

It is natural to expect that the probabilities of typical fluctuations and of
large deviations are either both enhanced or both reduced, in comparison with
the equilibrium system having the same $\bar\rho$, but this is not known to
be true in general.  In fact, our work here shows that small (typical)
fluctuations cannot in general be computed from the LDF. 

One can note that our expressions for the LDF (\ref{result}) and
(\ref{result3}) do not depend on the asymmetry parameter $0 \leq q < 1$. 
These expressions, however, are not valid at $q=1$, and they do not reproduce
the symmetric exclusion result: the limits $q \to 1$ and $N \to \infty$ do
not commute.  It would be interesting to analyze the case of large $N$ with
$1-q=O(N^{-1})$, in order to interpolate between the symmetric case and the
asymmetric case. 

It would also be desirable to have a physical understanding of our
additivity formulae (\ref{additivity-asym1}) and (\ref{additivity-asym2}),
and to see how our results could be generalized to more complicated
non-equilibrium steady states. 

Lastly, from our knowledge of the LDF in the SNS steady state, one could try
to determine how a given (unlikely) profile was produced dynamically out of
the nonequilibrium steady state.  This has been done for the symmetric
exclusion process on a circle in \cite{KOV}, and recently for the open
system in \cite{BDGJL,BDGJL2}, where the LDF was given in terms of a time
integral over a trajectory taking the system, via a ``reversed dynamics,''
from a typical SNS configuration to the profile $\rho(x)$.  A dynamic LDF for
the ASEP on the circle was studied in \cite{J}.

\section*{Acknowledgments}

 We thank T.~Bodineau, G.~Giacomin, and J.~M. Ortiz de Z\'arate for helpful
discussions.  The work of J.~L.~Lebowitz was supported by NSF Grant
DMR--9813268, AFOSR Grant F49620/0154, DIMACS and its supporting agencies,
and NATO Grant PST.CLG.976552.  J.L.L.  acknowledges the hospitality of the
Institut Henri Poincar\'e, and B.D. and J.L.L that of the Institute for
Advanced Study, where a part of this work was done.

\appendix

 \section{The concave envelope construction\label{con_env}}

In this appendix we justify the construction (\ref{CE},\ref{COCE}) of the
optimizing function $F_\rho$ for the supremum in (\ref{result}).  Recall that
we are given a density profile $\rho(x)$ which is
defined for $a\le x\le b$ and satisfies
$0\le\rho(x)\le1$ for all $x$,
and reservoir densities $\rho_a$ and $\rho_b$ which satisfy
$1\ge\rho_a>\rho_b\ge0$.  Let $H_\rho(x)=\int_a^x(1-\rho(y))\,dy$, so that
$G_\rho$ is the concave envelope of $H_\rho$, and recall 
that $F_\rho$ is obtained by cutting off
$G'_\rho(x)$ at $\rho_a$ and $\rho_b$ (see (\ref{COCE})).  Then (\ref{result})
may be written as
 \begin{eqnarray}
 \label{start}
   {\cal F}_{[a,b]} (\{\rho(x)\};\rho_a,\rho_b)
   &=& - (b-a) K(\rho_a,\rho_b)  \nonumber\\
  &&\hskip-115pt + \int_a^b dx\, [\rho(x)\log\rho(x)
     +(1-\rho(x))\log(1-\rho(x))]
 + \sup_{F(x)} B_\rho(F)\;,
  \end{eqnarray}
where
 \begin{equation}
  B_\rho(F)
  = \int_a^b dx\, \left[(1-\rho(x))\log{F(x)\over1-F(x)}
    +\log (1-F(x))\right]
  \end{equation}
 and the supremum in (\ref{start}) is over monotone nonincreasing functions
$F(x)$ satisfying $F(a)=\rho_a$, $F(b)=\rho_b$.  Now since $H_\rho(x)\le
G_\rho(x)$ for all $x$ and $\log[F(x)/1-F(x)]$ is decreasing, we obtain by
integration by parts, noting that $H(a)=G(a)$ and $H(b)=G(b)$, that
for any $F(x)$ in this class,
 \begin{eqnarray}
  \int_a^b dx\, \bigl[(1-\rho(x))-G'_\rho(x))\bigr]\,
         \log{F(x)\over1-F(x)} &&\nonumber\\
  && \hskip-140pt =\,  - \int_a^b \bigl[H_\rho(x))-G_\rho(x)\bigr]
   \,d\left[\log{F(x)\over1-F(x)}\right]\le0, 
  \end{eqnarray}
 so that 
 \begin{equation}
 \label{inter}
   B_\rho(F) \le\int_a^b dx \biggl[G'_\rho(x)\log{F(x)\over1-F(x)}
     + \log (1-F(x))\biggr].
  \end{equation}
 Since the integrand in (\ref{inter}) is pointwise concave in $F(x)$, with a
maximum at $F(x)=G'_\rho(x)$, the integral is, for the functions $F(x)$
satisfying $\rho_a\ge F(x)\ge\rho)b$, bounded above by its value at
$F(x)=F_\rho(x)$.  Thus
 \begin{eqnarray}
 \label{end}
   B_\rho(F) 
  &\le& \int_a^b dx\, \biggl[G'_\rho(x)\log{F_\rho(x)\over 1- F_\rho(x) } +
    \log (1- F_\rho(x)) \biggr] \nonumber\\
  &=& \int_a^b dx\, \biggl[(1-\rho(x))\log{F_\rho(x)\over 1- F_\rho(x) } +
    \log (1- F_\rho(x)) \biggr]\nonumber\\
  &=&B_\rho(F_\rho)\;;
   \end{eqnarray}
 here the first equality is obtained by noting that if $(c,d)$ is a maximal
interval on which $G_\rho(x)\ne H_\rho(x)$ then (i)~$G'_\rho(x)$ and hence
$F_\rho(x)$ are constant on this
interval and (ii)~$\int_c^dG'_\rho(x)\,dx=\int_c^d(1-\rho(x))\,dx$.  Equation
(\ref{end}) shows that the supremum in (\ref{start}) is achieved by
$F(x)=F_\rho(x)$. 

 \section{Optimal profile under a constraint\label{opt_constraint}}
 
 Let $\rho(x)$ be an optimal system profile 
for a fixed mean density $r$, that is, a profile which minimizes
$\F_{[a,b]}(\{\rho(x)\};\rho_a,\rho_b)$ under the constraint
 \begin{equation}\label{constraint}
\int_a^b \rho(x)\,dx = r(b-a).
 \end{equation}
 In this section we show that $\rho(x)$ is uniquely determined, and derive
its form. 

{\bf Case 1.  $\rho_a>\rho_b$}.  Let $G_\rho$ and $F_\rho$ be defined by
(\ref{CE}) and (\ref{COCE}), so that
 \begin{equation}
 \label{result_app}
 \F_{[a,b]}(\{\rho(x)\};\rho_a,\rho_b)
   =  \int_a^b h(\rho(x),F_\rho(x);\bar\rho)\,dx,
 \end{equation}
 where $h(r,f;\bar\rho)$ is defined in (\ref{hdef}) and $\bar\rho$ is
determined from $\rho_a,\rho_b$ as in Figure~1. 

 We first show that $\rho(x)$ must be monotone nondecreasing, that is, that
$\rho(x)=1-G'_\rho(x)$ (almost everywhere).  For otherwise
there will exist some interval $[c,d]\subset[a,b]$
satisfying
 \begin{equation}
 G_\rho(c)=\int_a^c(1-\rho(y))\,dy,\qquad
 G_\rho(d)=\int_a^d(1-\rho(y))\,dy,
 \end{equation}
 and
 \begin{equation}
 G_\rho(x)>\int_a^x(1-\rho(y))\,dy, 
      \qquad\hbox{for $c\le x\le d$.}
 \end{equation}
  Then for $c\le x\le d$, $G'_\rho(x)=1-\lambda$, where 
 \begin{equation}
 \lambda={1\over c-d}\int_c^d \rho(y)\,dy, 
 \end{equation}
  and if $\rho^*$ is defined by 
 \begin{equation}
 \rho^*(x)= \left\{
      \begin{array}{ll}
        \lambda ,& \mbox{if $c<x<d$,}  \\
        \rho(x),& \mbox{otherwise,}
      \end{array} \right.
 \end{equation}
 then $\rho^*$ satisfies (\ref{constraint}) and from the strict convexity of
$h(\rho,F;\bar\rho)$ in $\rho$ and the fact that $F_\rho$ is constant on $[c,d]$ (with
value $\lambda$, $\rho_a$, or $\rho_b$) it follows that
 \begin{equation}
  (c-d)h(\lambda,F_{\rho};\bar\rho)
  < \int_c^d h(\rho(x),F_{\rho}(x);\bar\rho)\,dx 
 \end{equation}
 and hence that  $\F_{[a,b]}(\{\rho^*(x)\};\rho_a,\rho_b) <
\F_{[a,b]}(\{\rho(x)\};\rho_a,\rho_b)$, contradicting the definition of
$\rho(x)$. 
 
Since $\rho(x)=1-G_\rho'(x)$ we have from (\ref{COCE}) that $F_\rho(x)$ is a
local function of $\rho(x)$ taking value $\rho_a$ if $1-\rho(x)>\rho_a$,
$\rho_b$ if $1-\rho(x)<\rho_b$, and $1-\rho(x)$ otherwise.  Then one can
check that the integrand $h(\rho(x),F_\rho(x);\bar\rho)$ in
(\ref{result_app}) is pointwise convex in $\rho(x)$.  Thus since $\rho(x)$
satisfies (\ref{constraint}),
$(b-a)h(r,1-r,\bar\rho)\le\F_{[a,b]}(\{\rho(x)\};\rho_a,\rho_b)$, with
equality if and only if $\rho(x)$ is the constant function $\rho(x)=r$.

{\bf Case 2.  $\rho_a<\rho_b$}.  Let $y_r$ be the minimizing
value in (\ref{result3}) corresponding to the  optimal profile
$\rho(x)$; it is clear that $\rho(x)$ must be constant on the intervals
$[a,y_r]$ and $[y_r,b]$, so that 
 \begin{eqnarray}
  \label{optimal_shock}
 \F_{[a,b]}(\{\rho(x)\};\rho_a,\rho_b)\hskip-60pt\\
   &=& (y_r-a)h(r_a,\rho_a;\bar\rho)+(b-y_r)h(r_b,\rho_b;\bar\rho)
 \nonumber
 \end{eqnarray}
 for some values $r_a,r_b$ satisfying
 \begin{equation}
 \label{constraint_shock}
(y_r-a)r_a+(b-y_r)r_b = (b-a)r. 
 \end{equation}
 Minimizing (\ref{optimal_shock}) over $a\le y_r\le b$ and
(\ref{constraint_shock}) leads to $y_r=b$, $r_a=r$ if $r<1-\rho_b$ and
$y_r=a$, $r_b=r$ if $r>1-\rho_a$, while if $1-\rho_b\le r\le 1-\rho_a$, $y_r$
is given by (\ref{y_crit})  and $r_a=1-\rho_b$, $r_b=1-\rho_a$.

\end{document}

%% file: phplane.tex
\hbox to \hsize{\hss
 \beginpicture
 \setcoordinatesystem units <3.0truein,3.0truein> point at 0 0
 \setplotarea x from 0.0 to  1.0, y from 0.0 to  1.0
 \plotsymbolspacing=0.1pt
 \setlinear
 \plot  0 0 1 0 1 1 0 1 0 0 /
 \arrow <10pt> [0.2,0.6] from 0.2 1.03 to 0.15 0.85
 \put {\dlsbig S} [b] at 0.22 1.04
 \setplotsymbol ({.})
 \setdashes
 \plot 0.004 0.004 0.996 0.996 /
 \setsolid
 \plot 0.998 0.5 0.5 0.5 0.5 0.002 /
 \setplotsymbol ({\bigsym .})
 \plot 0.005 0.995 0.5 0.5 /
% \plot 0.005 0.995 0.21 0.79 /
% \plot 0.29 0.71 0.5 0.5 /
% \put {\dlsbig F} [b] at 0.25 0.75
 \put {\dlsbig A\lower5pt\hbox{\dlsmed1}} at 0.2 0.55
 \put {$\rhob=\rho_a$} at 0.2 0.42
 \put {{\dlsbig A\lower5pt\hbox{\dlsmed2}}} at 0.37 0.25
 \put {$\rhob=\rho_a$} at 0.37 0.12
 \put {\dlsbig B\lower5pt\hbox{\dlsmed1}} at 0.5 0.8
 \put {$\rhob=\rho_b$} at 0.5 0.70
 \put {{\dlsbig B\hskip1pt\lower5pt\hbox{\dlsmed2}}} at 0.84 0.7
 \put {$\rhob=\rho_b$} at 0.84 0.60
 \put {\dlsbig C} at 0.75 0.3
 \put {$\rhob={1\over2}$} at 0.75 0.17
 \put {0} [t] at 0 -0.03
 \put {1} [t] at 1 -0.03
 \put {0} [r] at -0.03 0
 \put {1} [r] at -0.03 1
 \put {$\rho_a$} [t] at 0.6 -0.06
 \put {$\rho_b$} [t] at -0.06 0.6
 \endpicture
 \hss}

 \bigskip
 
\caption{The phase diagram of the open ASEP\label{fig1}}